\documentclass[12pt,preprint]{aastex}
\usepackage{lscape}

\newcommand{\bdv}[1]{\mbox{\boldmath$#1$}}
\def\E{{\rm E}}
\def\bpi{{\bdv{\pi}}}
\def\bvt{{\bdv{\tilde v}}}
\def\ret{{\tilde r_\E}}
\def\rehat{{\hat r_\E}}
\def\rel{{\rm rel}}
\def\pirel{\pi_{\rm rel}}
\def\au{{\rm AU}}
\def\mbr{{M_{\rm brk}}}
\def\bmu{{\bdv{\mu}_{\rm rel}}}
\def\kms{{\rm km}\,{\rm s}^{-1}}
\def\masyr{{\rm mas\,yr^{-1}}}
\def\dchi{\Delta\chi^2}
\def\dchix{\Delta\chi^2_{\rm xallarap}}
\def\dchip{\Delta\chi^2_{\rm parallax}}
\def\Rmacho{$R_{\rm MACHO}$ }
\def\Bmacho{$B_{\rm MACHO}$ }
\def\Reros{$R_{\rm EROS}$ }
\def\Beros{$B_{\rm EROS}$ }

\begin{document}
\title{
Systematic Analysis of 22 Microlensing Parallax Candidates
}

\author{Shawn Poindexter\altaffilmark{1},
	Cristina Afonso\altaffilmark{2},
	David P. Bennett\altaffilmark{3},
	Jean-Francois Glicenstein\altaffilmark{4},
        Andrew Gould\altaffilmark{1},
	Micha{\l} K. Szyma\'nski\altaffilmark{5}, and
	Andrzej Udalski\altaffilmark{5}
}
\altaffiltext{1}{Department of Astronomy, Ohio State University, 
140 West 18th Avenue, Columbus, OH 43210, USA,
(sdp,gould)@astronomy.ohio-state.edu}
\altaffiltext{2}{Max-Planck fuer Astronomie, Koenigstuhl 17, 69117 Heidelberg,
Germany, afonso@mpia-hd.mpg.de}
\altaffiltext{3}{Department of Physics, Notre Dame University,
Notre Dame, IN 46566, USA, bennett@nd.edu}
\altaffiltext{4}{DAPNIA-SPP, CE-Saclay, F-91191 Gif-sur-Yvette, France,
glicens@hep.saclay.cea.fr}
\altaffiltext{5}{Warsaw University Observatory, Al.~Ujazdowskie 4,
00-478 Warszawa, Poland,\break
(udalski,msz)@astrouw.edu.pl}


\begin{abstract}

We attempt to identify all microlensing parallax events
for which the parallax fit improves
$\dchi>100$ relative to a standard microlensing model.
We outline a procedure to identify three types of discrete degeneracies 
(including a new one that we dub the ``ecliptic degeneracy'')
and find many new degenerate solutions in 16 previously published and 6
unpublished events.
Only four events have one unique solution and the other
18 events have a total of 44 solutions.
Our sample includes three previously identified black-hole (BH)
candidates.  We consider the newly discovered degenerate solutions and
determine the relative likelihood that each of these is a BH.
We find the lens
of event MACHO-99-BLG-22 is a strong BH candidate (78\%), event
MACHO-96-BLG-5 is a marginal BH candidate (37\%), and MACHO-98-BLG-6 is a weak
BH candidate (2.2\%).
The lens of event OGLE-2003-BLG-84 may be a Jupiter-mass free-floating planet
candidate based on a weak $3\sigma$ detection of finite-source effects.
We find that event MACHO-179-A is a brown dwarf candidate within $\sim100$ pc
of the Sun, mostly due to its very
small projected Einstein radius, $\ret = 0.23\pm0.05\,\au$.
As expected, these microlensing parallax events are biased
toward lenses that are heavier and closer than average.  These events
were examined for xallarap (or binary-source motion), which can mimic
parallax.  We find that 23\% of these events are strongly affected
by xallarap.

\end{abstract}


\section{Introduction
\label{sec:intro}}

	Since microlens parallaxes were first predicted
\citep{gould92} and observed \citep{alcock95}, about 20 parallax
events have been reported in the literature.  The microlens parallax,
$\pi_{\rm E}$, expresses the size of the Earth's orbit (1 AU) relative 
the Einstein radius of the microlensing event projected onto the
observer plane $(\tilde r_{\rm E})$.  That is, 
$\pi_{\rm E}= {\rm AU}/\tilde r_{\rm E}$.  It is related to the lens
mass $M$ and the lens-source relative parallax $\pi_{\rm rel}$ by
\begin{equation}
\pi_\E = \sqrt{\frac{\pirel}{\kappa M}},
\qquad \kappa = {4 G\over c^2\,\rm AU}\sim 8.1{{\rm mas}\over M_\sun},
\label{eqn:pie}
\end{equation}
The microlens parallax is determined by modeling the
distortion in the microlensing light curve, relative to the
standard \citep{pac86} shape, that is generated
by the deviation of the Earth's motion from a straight line.
Other things being equal, the bigger the parallax, the greater the
distortion.

	Microlens parallaxes can help break the classic timescale degeneracy
in microlensing events and, because of this, they have a wide range of
potential applications.  For most events, the only parameter that one can
measure that gives any information about 
the underlying physical characteristics of
the lens is the Einstein timescale $t_{\rm E}$.  This is related to
$M$, $\pi_{\rm rel}$ and the lens-source relative proper motion $\mu_{\rm rel}$
by
\begin{equation}
t_{\rm E} = {\theta_{\rm E}\over \mu_{\rm rel}},
\qquad \theta_{\rm E} = \sqrt{\kappa M \pi_{\rm rel}},
\label{eqn:tethetae}
\end{equation}
where $\theta_{\rm E}$ is the angular Einstein radius.
Hence, by measuring $\pi_{\rm E}$, one eliminates the uncertainties
arising from the unknown relative proper motion and so obtains a direct
relation between $M$ and $\pi_{\rm rel}$ through equation (\ref{eqn:pie}).
\citet{alcock95} used this relation to place constraints on the lens mass and 
distance for the very first parallax event, and the same principle has
been applied for many subsequent events.  Indeed, \citet{bennett02},
\citet{mao02}, and
\citet{agol02} used parallax measurements to argue that the lenses of
three microlensing events were black-hole
candidates.  Moreover, \citet{hangould95} showed that an ensemble of parallax
events could be used to constrain the mass function of the lenses

	If both $\pi_{\rm E}$ and $\theta_{\rm E}$ are measured, one can
determine both $M=\theta_{\rm E}/\kappa\pi_{\rm E}$ and 
$\pi_{\rm rel} = \pi_{\rm E}\theta_{\rm E}$ \citep{gould92}.  To date,
this has been done fairly accurately for three events
\citep{an02,gould04,gba04,kubas},
and more crudely for one other \citep{jiang04}.

	However, microlens parallaxes are subject to their own degeneracies.
To understand these, it is necessary to recognize that the microlens
parallax is actually a vector $\bpi_{\rm E}$, whose magnitude is given
by $\pi_{\rm E}$ and whose direction is that of the lens motion relative
to the source.
For relatively short microlensing events, $t_{\rm E}\la {\rm yr}/2\pi$,
the Earth's acceleration may be approximated as roughly constant during
the event.  \citet{gmb} pointed out that this would give rise to
a strong asymmetry in the light curve but otherwise weak effects.
Since the magnitude of this asymmetry is proportional to 
$\pi_{{\rm E},\parallel}$ (the component of $\bpi_{\rm E}$ parallel
to the Earth's acceleration), one could potentially measure this 
component of the vector parallax in such short events, but not the
other ($\pi_{{\rm E},\perp}$).
Such essentially 1-dimensional parallax measurements have
subsequently been made for three events \citep{park04,ghosh04,jiang04}.

	On closer examination of this limit, \citet{smith03} found
that even if the Earth's motion were approximated as uniformly accelerating,
one could in principle determine both components of $\bpi_{\rm E}$.
However, they noted three distinct types of degeneracies.  The first
is a two-fold discrete degeneracy, essentially whether the lens
passes the source on its right or left.  This degeneracy, which gives rise to 
relatively small changes in $\bpi_{\rm E}$ as well as other event parameters,
is known at the ``constant-acceleration'' degeneracy.  \citet{smith03}
showed that even though the Earth's acceleration is not perfectly uniform,
the constant-acceleration degeneracy did affect a number of archival events.
The second degeneracy is a continuous one between $\pi_{{\rm E},\perp}$
and blending of the light curve by light from an additional, unmagnified
source.  Since both of these give rise to light-curve distortions that
are even about the peak, they can be mistaken for each other unless the
data are of sufficiently high quality.  Indeed, the source flux, the
background flux, the event timescale, and the event impact parameter
all affect the light curve symmetrically about the peak, and disentangling
these effects is a classic problem in the interpretation of microlensing
events.  Adding $\pi_{{\rm E},\perp}$ to the mix simply exacerbates
an already difficult situation.  This second degeneracy therefore leads
to 1-dimensional parallax measurements and is essentially the same as the
one identified by \citet{gmb}.  Finally, \citet{smith03} noted that
if the source were accelerating (due to a binary companion) it could
produce exactly the same light-curve distortion as the Earth's acceleration.
This degeneracy is especially severe if both accelerations
can be approximated as
uniform but, at least in principle, the binary orbit of the source
could mimic with infinite precision the orbital parameters of the Earth.
In practice, however, one would not expect a binary source to have exactly
the same orbital parameters as the Earth.  Hence, if the data are
of sufficient quality to measure these parameters, and if the measured
parameters are inconsistent
with those of the Earth, this would imply that the light curve is affected
by the acceleration of the source rather than the Earth.  Conversely,
if the fit parameters closely mimic those of the Earth, one can infer
that the light-curve distortions are most likely due to parallax.

In principle, acceleration of the lens could also produce light-curve
distortions that mimic parallax.  However, in most cases, if the lens
has a companion with a period close to a year, the light curve will
be far more severely affected by the binarity of the lens than by
the accelerated motion of one of its components.

	To mathematically analyze the constant-acceleration degeneracy,
\citet{smith03} Taylor-expanded the lens-source separation to fourth order
in time.  \citet{gould04} adopted this same approach, but included the
Earth's jerk as well as its acceleration.  Surprisingly, this addition
led to the identification a new, so-called ``jerk-parallax'', degeneracy.
In contrast to the constant-acceleration degeneracy, the jerk-parallax
degeneracy can lead to radically different estimates of $\bpi_{\rm E}$,
both in its magnitude and direction.  Indeed, \citet{gould04} was
specifically motivated to make this analysis by his empirical discovery
of two very different parallax solutions for the event MACHO-LMC-5.  These
solutions led to very different mass and distance estimates for the lens,
an ambiguity that was finally resolved by a direct (trigonometric)
parallax measurement
by \citet{dck04}.  \citet{gould04} showed that his formalism, although
idealized, predicted the second solution, given the first, extremely well.
Moreover, \citet{park04} showed that this formalism also predicted
the location of a second solution in another event, giving confidence
that one could indeed find additional solutions analytically if they
existed.

This confidence is important because the previous practice was to
search for additional parallax solutions by brute-force, i.e., by starting
with seed solutions at many different places in parameter space and then
moving downhill on the $\chi^2$ surface until reaching a local minimum.
Although such an approach can yield additional solutions, there is
no guarantee that it will find them if they exist.  Indeed, we
will show in this paper that this brute-force approach did in fact fail
to find additional parallax solutions for events that were previously
analyzed.

A crucial component of the \citet{gould04} approach is to work in the
geocentric frame, i.e., the Galilean frame that is at rest with
respect to the Earth at the peak of the event.  In this frame, all
the parameters characterizing the event (except the parallax) are
approximately the same for all solutions (up to the left-right ambiguity
mentioned above).  This means that searches, even when they are not
guided by analytic insight, are much more likely to be successful.

	Here we undertake a systematic study of all microlensing
events with parallax signatures that are detected with ``good confidence'', 
which we define as a $\dchi>100$ improvement relative to the standard
(non-parallax) solution.  We seek to identify all parallax solutions
(if there are more than one) and to determine whether the event is
better fit by ``xallarap'' (binary-source motion) rather than parallax.
We aim to achieve several interrelated goals.

First, at present it is unknown how severely parallax events are affected
by discrete degeneracies.  These have been systematically searched for
in only four events, MACHO-LMC-5 \citep{gould04}, MOA-2003-BLG-37
\citep{park04}, OGLE-2003-BLG-175/MOA-2003-BLG-45 \citep{ghosh04},
and OGLE-2003-BLG-238 \citep{jiang04}.  None of these events
satisfy $\dchi>100$ and so none are in the present study.
In principle, therefore, it is possible that the events that do
satisfy this criterion are not seriously affected by degeneracies.

Second, by examining a large ensemble of events, we seek to identify
patterns in the event degeneracies.  In fact, we find a
new class of degeneracy, which we dub the ``ecliptic''  degeneracy.
As noted by \citet{jiang04}, events lying exactly on the ecliptic
will be subject to an exact two-fold degeneracy.  We show that
this degeneracy is a combination of the previously identified
constant-acceleration and jerk-parallax degeneracies.  Moreover,
we show that, since events seen toward the Galactic bulge are
generally quite near the ecliptic, this degeneracy indeed affects
many of these events.

Third, we seek to reanalyze the three black-hole candidates, taking
account of any new degenerate solutions that we find.  In particular,
we calculate the relative likelihood that these events are due to
main-sequence stars, white dwarfs, neutron stars, and black-holes.
We find that one of the three is a strong candidate, one is
a very weak candidate, and one is a plausible, but not strong,
candidate.

Fourth, we analyze the likelihood distributions of all events with
respect to mass and distance.
As has been frequently noted, parallax events are expected to be
biased toward lenses that are heavier, closer, and slower than average.
We confirm this expectation.

Fifth, we analyze all events for xallarap.  Of events with apparent
parallax signatures, for what fraction is this actually due to
xallarap?  Of course, the sample is somewhat biased since xallarap
due to short-period binaries would not mimic parallax effects very
well and so might not make our cut.  Nevertheless, we expect that
most xallarap events with recognizable signals will be in our sample,
giving us a good probe of the relative frequency of parallax and
xallarap events.  We find that $23\%$ of our parallax-candidate sample
is strongly affected by xallarap.

Finally, we check our sample for any intrinsically interesting events
other than the black-hole candidates that have been previously
identified. Indeed, we find a brown-dwarf candidate, which probably lies
within $\sim$100 pc of the Sun.
We also find a candidate free-floating planet of mass $M \sim 10^{-3} M_\sun$.
In all we analyze 22 events, of which 16 are taken from the literature
and 6 are previously unanalyzed events from the OGLE-III database.

\section{Microlensing Parallax Sample
\label{sec:sample}}

	We attempt to identify all point-lens microlensing
parallax events with improvements of
$\dchi > 100$ relative to a standard
(non-parallax) microlensing model\footnote{\citet{an02} have measured
the path of a binary event, EROS BLG-2000-5. While they did not directly
determine $\dchi$, their error estimate for $\bpi_\E$ is $3\%$, indicating
a $30\sigma$ detection, which would appear to meet our $\dchi$ criterion.
In addition, \citet{kubas} measured the $\pi_\E$ for the binary event
OGLE-2002-BLG-069 to a precision of $16\%$ (after rescaling
errors) corresponding to $6\sigma$, which probably would not
meet our $\Delta\chi^2$ criterion.  However, in any case, binary
parallax events are much more complicated than point-lens
events, so we exclude them from considerations here.
}.
To do so, we first search the literature for
parallax events with published models.
We identify 16 such events, including
10 discovered by MACHO\footnote{\url{data available at
http://www.macho.mcmaster.ca}}
\citep{alcock93,bennett02},
4 discovered by OGLE\footnote{\url{data available at
http://www.massey.ac.nz/~iabond/alert/alert.html}}
\citep{udalski},
1 discovered by MOA\footnote{\url{data available at
http://ogle.astrouw.edu.pl/ogle3/ews/ews.html}}
\citep{bond01}, and
1 discovered by EROS
\citep{afonso}.
Some events discovered by one collaboration
were simultaneously monitored by another, and
we attempt to include these data sets in
our analysis.
Some parallax events in the literature such as
MACHO-LMC-5 \citep{alcock97b,alcock01,gould04} did not make our
$\dchi > 100$ cut.

	\citet{bennett02} identified nine MACHO
parallax events that made our sample, but did not publish their analysis
of three of these because of their more severe $\dchi$ threshold.
We find that one event that they did analyze, MACHO-99-BLG-8, has
$\dchi = 56.51$. However, we also find that the standard
solution has severe negative blending (an unrealistic result), and that
when the background is constrained to zero for the standard solution,
$\dchi = 1190.27$.  Similarly, we find that EROS-BLG-29 has
$\dchi=41.44$, but also with severe negative blending.  When
the blending is constrained to zero, $\dchi=129.46$.

	Next we search for unpublished parallax events
from on-line data obtained by the OGLE-III \citep{udalski03}
and MOA \citep{bond01} collaborations among events
identified up though the 2003 seasons.
These events are visually examined if the geocentric
timescale is $t_\E \geq 60$ days.
We then fit plausible parallax candidates with both
standard and parallax microlensing models and adopt
those with $\dchi \ge 100$. We find 6 such events,
all from the the OGLE-III database.
The microlensing events are listed in Table 1.
Most of the headings of
this table are self-evident. The baseline magnitude of the event
$m_{\rm source}$ is a calibrated Johnson/Cousins $R$ for MACHO events, a
calibrated Cousins $I$ for OGLE-II events (1999 and 2000), an
approximately calibrated Cousins $I$ for OGLE-III events (2002 and 2003),
an approximately calibrated Cousins $I$ for MOA-2000-BLG-11, and a
calibrated Cousins I for EROS-BLG-29.
The one exception to this is event MACHO-99-BLG-22 in which the Cousins
$I$ band is given because the MACHO R band is corrupted.
The column ``points'' give the total number
of original data points and the number removed. The renormalization factors
are given in an order determined as follows.
For each event, the band(s) of the collaboration whose name is attached to the
event are given first.
For MACHO, these are two bands
$R_{\rm MACHO}$ and $B_{\rm MACHO}$, and these are given in this order.
When there are additional bands, these are given in the order they are
mentioned in \S~\ref{sec:solutions}. For event MACHO-104-C we re-normalize
the errors separately for the peak and baseline (displayed on each side of
``/'', respectively).

\section{Parallax Solutions
\label{sec:parsolutions}}

	We first fit each event to a standard \citet{pac86} curve,
\begin{equation}
F_i(t) = f_{s,i} A[u(t)] + f_{b,i},\qquad A(u) = {u^2+2\over u\sqrt{u^2+4}},
\label{eqn:paceq}
\end{equation}
where $F_i$ is the observed flux at each of $n$ observatory/filter 
combinations, $f_{s,i}$ and $f_{b,i}$ are the source and background fluxes
for each of these combinations, and $A(u)$ is the magnification as a
function of $u$, the source-lens projected separation normalized to the
angular Einstein radius $\theta _\E$.  In the standard 
(i.e., non-parallax) model, the motions of the observer, source, and lens
are all assumed to be rectilinear, so $u$ is given simply by the Pythagorean
theorem,
\begin{equation}
u(t) = \sqrt{\beta^2 +\tau^2},\qquad \tau = 
{t-t_0\over t_\E}, \quad \beta=u_0,
\label{eqn:pythag}
\end{equation}
where $t_0$ is the time of closest approach, $u_0 = u(t_0)$ is the
impact parameter, and $t_\E$ is the Einstein timescale. 
This procedure almost always converges, although for some events with very
distorted light curves the fit is rather poor. The two exceptions
to this are MACHO-99-BLG-1 and OGLE-2003-BLG-32, which have such
distorted light curves that $t_\E$ had to be held at a fixed value to
permit convergence.

	To search for parallax solutions, we adopt the geocentric
frame of \citet{gould04}, i.e., the Galilean frame that is coincident
with the position and velocity of the Earth at the peak of the
event.  In this frame, the predicted fluxes are given by equations
(\ref{eqn:paceq}) and (\ref{eqn:pythag}), but with $\beta$ and
$\tau$ adjusted by
\begin{equation}
\tau = {t-t_0\over t_\E} +\delta\tau,\quad\beta = u_0 + \delta\beta, 
\label{eqn:betatau}
\end{equation}
where
\begin{equation}
(\delta\tau,\delta\beta) =
  (\bpi_\E \cdot \Delta{\bf s},\bpi_\E \times \Delta{\bf s}),
\label{eqn:deltabetatau}
\end{equation}
$\Delta\bf s$ is the positional offset
of the Sun projected onto the sky (and normalized
by an AU), 
and $\bpi_\E$ is a new set of two parameters, the ``vector microlens
parallax''.  

\subsection{Degeneracy Search}

	It is straightforward to find one parallax solution.  We
simply use the non-parallax solution (with $\bpi_\E=0$) as
a seed and search for a minimum in $\chi^2$.  Since the parameters
$t_0$, $u_0$, and $t_\E$ are very similar for the non-parallax
solution and for the parallax solutions in the geocentric frame, this
procedure always converges rapidly.  However, it leaves open the
question of whether there are other
parallax solutions that are degenerate with this initial one.
\citet{smith03} and \citet{gould04} identified two
types of degeneracy (respectively the ``constant-acceleration''
and the ``jerk-parallax'' degeneracies) to which events are subject
in the limit of weak parallax effects.
In the geocentric frame, the constant-acceleration degeneracy
is characterized by $u_0\rightarrow -u_0$,
with the other parameters changing very little.  
The ``jerk-parallax degeneracy'' sends
\begin{equation}
\pi_{\rm E,\parallel} \rightarrow \pi_{\rm E,\parallel},\qquad
\pi_{\rm E,\perp} \rightarrow -(\pi_{\rm E,\perp} + \pi_{j,\perp}),
\label{eqn:pieperp}
\end{equation}
where the parallel and perpendicular directions are defined by the
Sun's apparent acceleration at $t_0$ and
$\bpi_j$ is the ``jerk parallax''.  In this case also, the remaining
parameters change very little.  \citet{gould04} gives the exact formula
for $\bpi_j$, but in the approximation that the Earth's orbit is
circular, the perpendicular component is
\begin{equation}
\pi_{j,\perp} = -{4\over 3}{{\rm yr}\over 2\pi t_\E}
{\sin\beta_{\rm ec}\over
(\cos^2\psi\sin^2\beta_{\rm ec}+\sin^2\psi)^{3/2}},
\label{eqn:pijperp}
\end{equation}
where $\beta_{\rm ec}$ is the ecliptic latitude of the event
and $\psi$ is the phase of the peak of the event relative to
opposition.

\subsection{Constant-Acceleration Degeneracy
\label{sec:constaccel}}

	The constant acceleration degeneracy is the most common
and easiest to
identify, and we therefore look for it first.
It is obtained by using the first parallax solution
as a seed but with the sign of $u_0$ reversed. While
this always converges to a new solution, in five cases the
two solutions are not truly degenerate since one of these two
has a significantly worse $\chi^2$ or heavy negative blending.
If this potential degeneracy is not realized, then the
other possible degeneracies are not present either.

\subsection{Jerk-Parallax Degeneracy}

	Since the Galactic bulge lies close to the ecliptic, the
Sun's projected
apparent acceleration is generally parallel to the ecliptic, which
at the position of the Galactic bulge, lies along the East-West axis.
The vector-parallax components 
$\pi_{\E,\parallel}$ and $\pi_{\E,\perp}$ are therefore approximately
aligned with $\pi_{\E,E}$ and $\pi_{\E,N}$, the east and
north components of this vector.
Moreover, from equation (\ref{eqn:pijperp}) one finds that near the ecliptic,
$\pi_{j,\perp}\sim 0$.  Hence, equation (\ref{eqn:pieperp}) becomes
approximately,
\begin{equation}
\pi_{{\rm E},E} \rightarrow \pi_{{\rm E},E},\qquad
\pi_{{\rm E},N} \rightarrow -\pi_{{\rm E},N}.
\label{eqn:pieen}
\end{equation}
We therefore generally make this substitution in the original
solution to obtain a seed to search for the jerk-parallax degenerate
solution.  When this fails, we use the more exact formula of
\citet{gould04}. 
Three of the 22 events have at least 3 solutions.

	When the search for a third solution is successful, we reverse the
sign of $u_0$ in this solution to obtain a seed to search for a fourth
solution.  Only two of the 22 events have four distinct solutions.

\subsection{Ecliptic Degeneracy}

	As noted by \citet{jiang04}, events that lie exactly on the
ecliptic suffer a two-fold degeneracy.  Unlike the degeneracies
identified by \citet{smith03} and \citet{gould04}, which are
perturbative and so can be broken for events that are sufficiently
long or have sufficiently high-quality data, the ecliptic degeneracy
is exact to all orders.  Since the bulge lies near the ecliptic,
one expects this degeneracy to apply approximately to bulge events.
From simple geometric considerations, the exact ecliptic degeneracy
takes
\begin{equation}
u_0 \rightarrow -u_0,\qquad
\pi_{\rm E,\perp} \rightarrow -\pi_{\rm E,\perp}.
\label{eqn:ecdegen}
\end{equation}
Hence, toward the bulge, a good seed for the 
approximate ecliptic degeneracy can be obtained by the
substitutions, $u_0 \rightarrow -u_0$,
$\pi_{\rm E,N} \rightarrow -\pi_{\rm E,N}$.
While we did not, in fact, locate this degeneracy in this manner,
we find in retrospect that almost all of the events for which
there are at least two solutions do in fact suffer from the ecliptic
degeneracy and that it could have been found by the above substitution.
What occurred in practice is that when we reversed the sign
of $u_0$, but not $\pi_{\E,N}$, (see \S~\ref{sec:constaccel}),
the minimization procedure drove $\pi_{\E,N}$ to the opposite sign
anyway.

\subsection{Solutions
\label{sec:solutions}}

\begin{figure}
\plotone{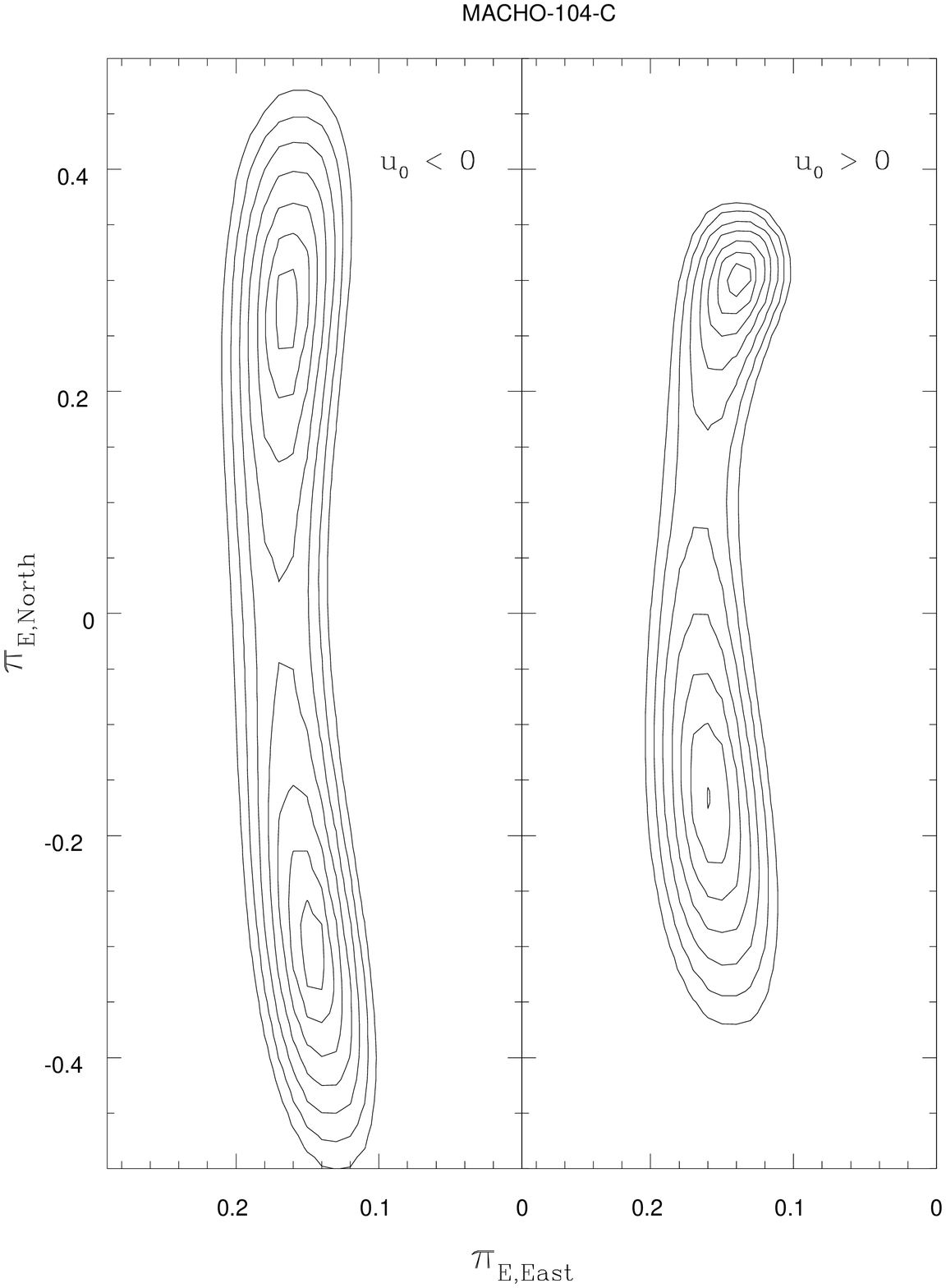}
\caption{$\dchi$ contour map in the $\bpi_\E-{\rm plane}$
for event MACHO-104-C.  The pairs of solutions in each panel is the
jerk-parallax
degeneracy. The solutions for positive $u_0$ and negative $u_0$ are at
similar $\bpi_\E$ because of the constant-acceleration degeneracy.
The ``ecliptic degeneracy'' identifies the upper (lower) solution in the
left panel with the lower (upper) solution in the right panel.
\label{fig:chi2-104c}}
\end{figure}

\begin{figure}
\plotone{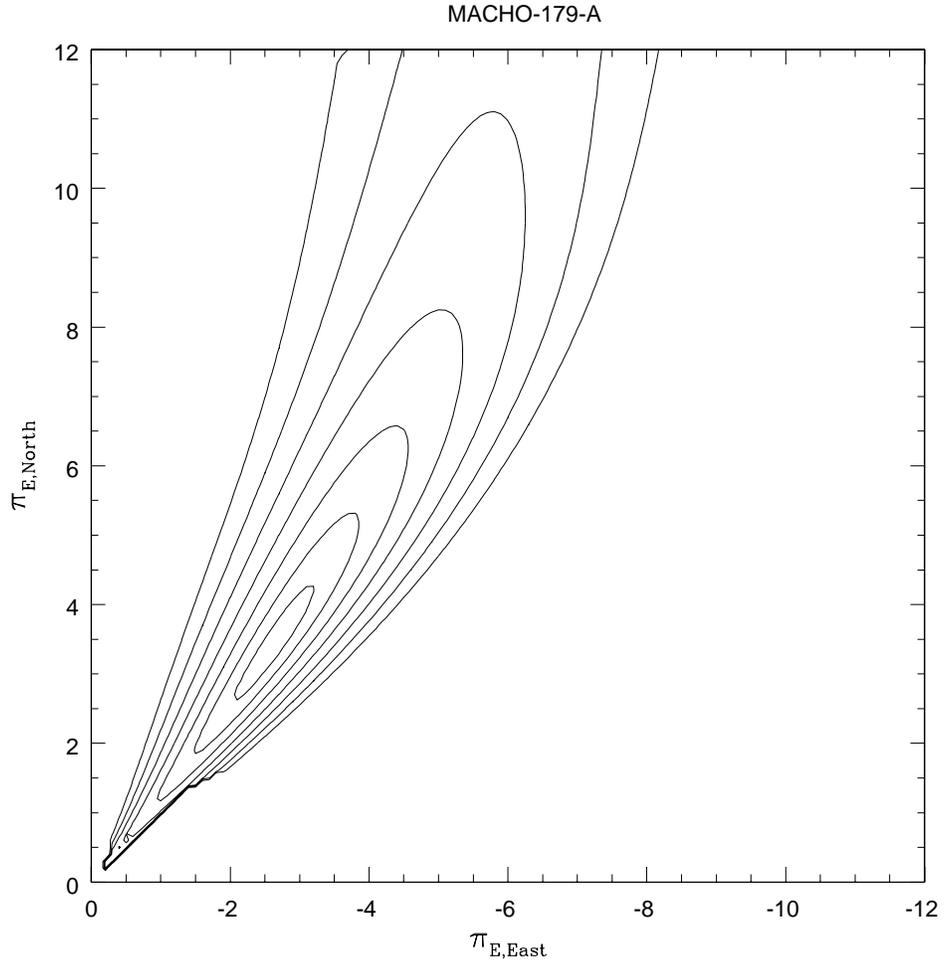}
\caption{$\dchi$ contour map in the $\bpi_\E-{\rm plane}$
for event MACHO-179-A.
\label{fig:chi2-179a}}
\end{figure}

	Once all the parallax solutions are found for a given event,
we focus our attention on the one with the lowest $\chi^2$.
We recursively remove outliers and rescale the errors so that
$\chi^2$ per degree of freedom (dof) is equal to unity.
We terminate this procedure when the largest outlier has
$\chi^2 < 14$.  This cleaned and renormalized data set is used to
evaluate all solutions.
A $\chi^2$ map in the $\bpi_\E$ plane is generated
for each solution to verify that all degenerate solutions
have been identified. This is shown for event MACHO-104-C in Figure
\ref{fig:chi2-104c}, and for event MACHO-179-A in Figure
\ref{fig:chi2-179a}.
Each solution is listed in Table 2.
Again, most of the table headings are self-evident.
The source magnitude is derived from the source flux for the band
for which the baseline magnitude (derived from $f_s+f_b$) is given
in Table 1. For MACHO events, this is actually a
combination of the two observed bands.
The column $\eta_b\equiv f_b/(f_s+f_b)$
gives the ratio of unlensed background-light flux to the total
baseline flux. The ``geocentric'' parameters are those obtained in the
fit. The ``heliocentric'' parameters are derived from these and
describe the event as it would be seen from the Sun. In particular,
$\theta$ is the angle of the lens-source relative motion, counterclockwise
(celestial) north through east.
An asterisk after the $\dchi$ indicates that the solution has the
background flux parameter ($f_{b,i}$ in equation [\ref{eqn:paceq}]) 
fixed to zero because the unconstrained solution (listed in the
preceding row) has an unrealistically negative blend.
(Blending can be slightly negative, while still remaining ``physical'',
because the ``sky'' in crowded bulge fields comes partly from a
mottled background of main-sequence stars.  If this background
happens to be lower at the source position than at neighboring
positions, $f_b$ will be slightly negative.  
See \citealt{park04} and \citealt{jiang04}.)
The column ``D'' gives the number of degenerate solutions.
Solutions in this paper are classified as ``degenerate'' if $\dchi<10$ and
the fit has realistic blending (unlike some of the parallax solutions of
EROS-BLG-29).
The degenerate solutions are in bold print in Table 2.
We now comment on individual events.

	Event MACHO-104-C, the first microlensing parallax event
ever discovered \citep{alcock95}, 
is one of two events in our sample
to have four degenerate parallax solutions.  In this case all 4 solutions
have $\dchi < 1$.
This is surprising considering that the parallax fit is well
constrained: the standard microlensing
model has $\dchi = 1647$ relative to the best
parallax solution. \citet{bennett02} found only
one solution.
Figure \ref{fig:chi2-104c} shows $\dchi$ contours
(1, 4, 9, 16, 25, 36, 49) in the $\bpi_\E-{\rm plane}$.

	Event MACHO-96-BLG-5 has been previously identified as a black-hole
candidate \citep{bennett02}.
We include the \Rmacho and \Bmacho bands along with
follow-up $R$ band observations from the MACHO/GMAN Project \citep{becker}.
We have excluded \Rmacho data from the 1999 season because the red CCD was
changed and this could create a systematic offset.  If we do include this
additional data, the best-fit solution increases its geocentric timescale from
$546\pm165$ days to $698\pm303$ days and the $\ret$ increases from
$16\pm5 \au$ to $21\pm9 \au$.  Additionally, the blending fraction
increases slightly.  This event is nearly at baseline and has no significant
slope so including it as a separate band gives no leverage on further
constraining the solution.
The two degenerate solutions differ in
velocity direction by $\sim 100\arcdeg$.  This difference affects the
mass estimate of the lens (see \S~\ref{sec:bh}).
This event suffers from the ecliptic degeneracy.

	Event MACHO-96-BLG-12 was initially found to have four degenerate
solutions.
After we rescaled the errors and removed 10 outliers, two of the solutions
merged leaving only 3 degenerate solutions.
We include the \Rmacho and \Bmacho bands along with
follow-up $R$ band observations from the MACHO/GMAN Project.
After adding 1275 EROS data points in the \Reros and \Beros bands,
we find the previous best solution becomes the worst, and the
previous second best solution becomes the best.  However, since the
entire range of $\dchi$ is only 3.25, such fluctuations are not
unexpected.
This event suffers from the ecliptic degeneracy.

	Event MACHO-98-BLG-6 has also been previously identified as a
black-hole candidate \citep{bennett02}. See \S~\ref{sec:bh}.
We include the \Rmacho and \Bmacho bands along with
follow-up $R$ band observations from the MACHO/GMAN and the
Microlensing Planet Search (MPS) Project \citep{rhie}.
This event suffers from the ecliptic degeneracy.

	Event MACHO-99-BLG-1 includes \Rmacho and \Bmacho bands along with
follow-up $R$ band observations from the MACHO/GMAN and 
MPS Projects.
This event suffers from the ecliptic degeneracy.

	Event MACHO-99-BLG-8 includes \Rmacho and \Bmacho bands along with
follow-up $R$ band observations from the MACHO/GMAN and MPS.
This event suffers from the ecliptic degeneracy.

	Event MACHO-179-A has only one solution, which has an unusually small
$\ret = 0.23$.
In \S~\ref{sec:BDcandidate} we show this small $\ret$ suggests
that the lens is a brown dwarf.
Figure \ref{fig:chi2-179a} shows $\dchi$ contours
(1, 4, 9, 16, 25, 36, 49) in the $\bpi_\E-{\rm plane}$.

	Event MACHO-95-BLG-27
includes follow-up $R$ and $B$ band observations from MACHO/GMAN and
University of Toronto Southern Observatory (UTSO) $R$ band
observations.
We find this event has four degenerate solutions. This relatively
high-magnification $(A_{\rm max} \sim 40)$ event shows no evidence for finite
source effects.

	Event MACHO-99-BLG-22 was discovered by MACHO then found by OGLE
as OGLE-1999-BUL-32 independently 2 months later.
We include data from both collaborations as well as EROS \Beros band,
MACHO/GMAN, and MPS data.
However, we exclude the MACHO R data because it is corrupted.
\citet{mao02} searched for degeneracies
but failed to find any. We find two highly degenerate solutions with
$\dchi=0.75$, another example of the ecliptic degeneracy. However, this
newly discovered solution is very similar to the previous one and has a very
small impact on the lens mass estimate in \S~\ref{sec:bh}.

	Event MOA-2000-BLG-11 suffers from the ecliptic degeneracy.

	Event OGLE-1999-BUL-19 has no degeneracies. \citet{smith02} also
searched for degeneracies and also found none.

Event OGLE-1999-CAR-01 has two solutions after reduction by the OGLE-III
pipeline.  When we first analyzed this event we had only the DoPhot photometry
which resulted in four degenerate solutions.  This shows how improved
reduction can break degeneracies.  This is
the only event we examined that is not near the ecliptic.
\citet{mao99} only identified one solution.

	Event OGLE-2000-BUL-43 does not have any degenerate solutions.
\citet{soszynski01} found two solutions with $\dchi=6.8$. When using
the same data set (which ends three days before the peak) we found the
same two solutions with $\dchi=7.15$. After including three seasons
of data after the peak, the degeneracy is broken.
This is an OGLE-II event, which was originally reduced with DoPhot
photometry.  However, our analysis is based on a re-reduction using
the OGLE-III image-subtraction pipeline. 

	We find sc33\_4505 has two solutions with $\dchi=8.23$.
\citet{smith03} searched for degeneracies of this event and
found a second solution with $\dchi=10.5$. The difference in
$\dchi$ derives from the fact that we rescaled our errors.
This is another example of the ecliptic degeneracy.

	Event OGLE-2002-BLG-100 is one of two events that has at least
two solutions but does {\it not} suffer from the ecliptic degeneracy.
This is not surprising since it is the farthest from the ecliptic of all
the events toward the bulge.

	Event OGLE-2002-BLG-334 has only one viable solution. The other
potential solution is ruled out by severe negative blending.

	Event OGLE-2002-BLG-61 suffers from the ecliptic degeneracy.

	Event OGLE-2003-BLG-188
was also monitored as MOA-2003-BLG-61. We include these
data in our analysis. It is one of two events with two or more
solutions that does {\it not} suffer from the ecliptic degeneracy.

	Event OGLE-2003-BLG-32 suffers from the ecliptic degeneracy.

\begin{figure}
\plotone{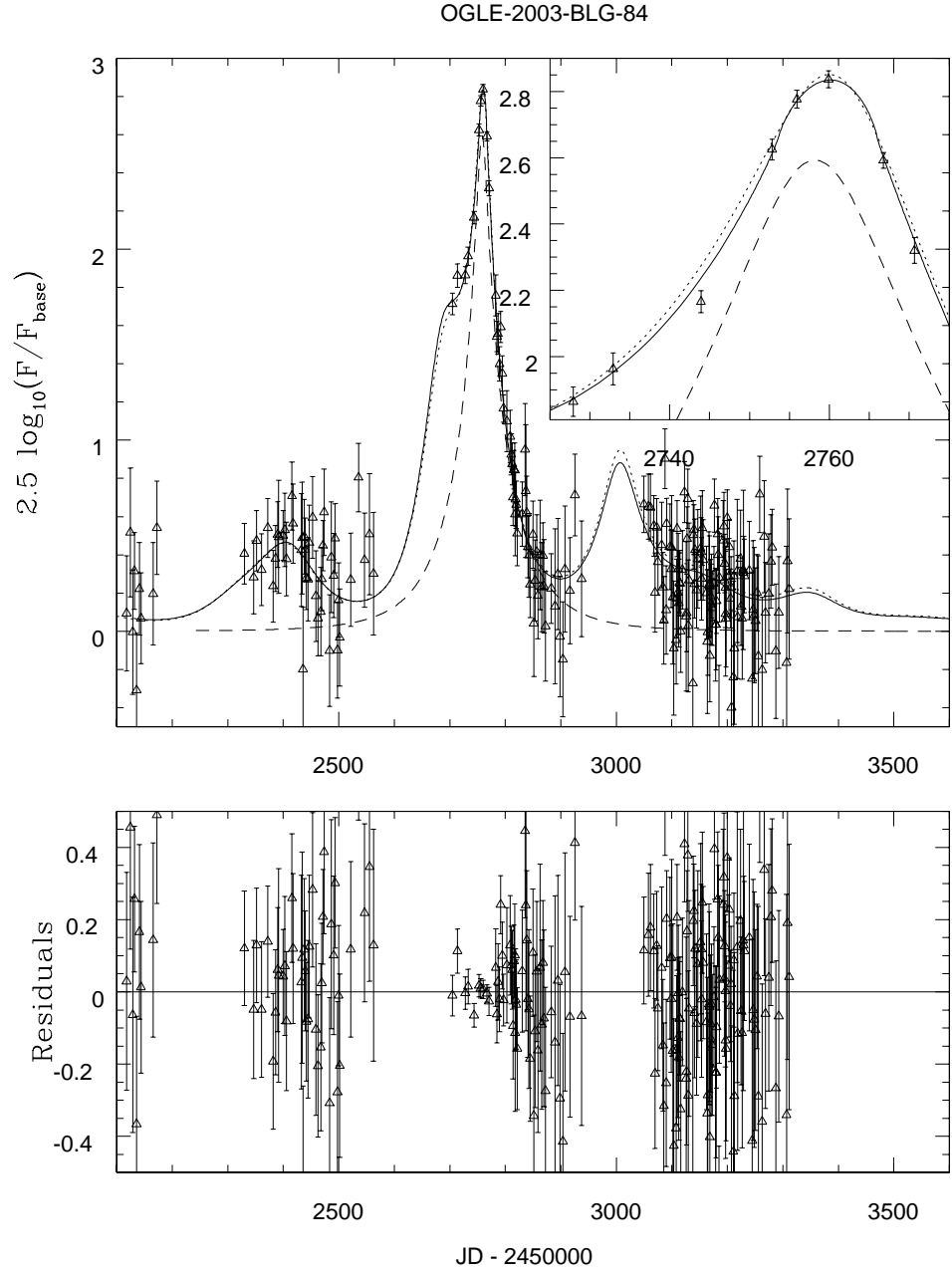}
\caption{Light curve of OGLE-2003-BLG-84 planetary candidate. The solid
curve is the best point-source parallax model with finite-source effects.
The dotted line is the best parallax model. The dashed line is the
non-parallax model. The upper right panel is an enlarged view of the light
curve peak. The bottom panel shows the residuals of the parallax model with
finite-source effects.
\label{fig:planetlc}}
\end{figure}

	Event OGLE-2003-BLG-84 is listed as a binary lens
in \citet{jaroszynski}. We include the 2004 data in our
analysis and find a hump in this previously unanalyzed year that 
is well-modeled by the parallax solution.
Moreover, as we discuss in \S~\ref{sec:xallarap}, the xallarap
fit reproduces the Earth's orbital parameters, which would be most
extraordinary if this were a binary event that just happened to be fit
by parallax. We infer that its complex structure (see Fig \ref{fig:planetlc})
is indeed due to parallax.
This event has two ``ecliptic'' degenerate solutions ($\dchi \le 7.85$).
Curiously one of these solutions predicts
a large spike in the light curve
between the 2003 and 2004 observing season, although the better one does not.
As we discuss in \S~\ref{sec:planet}, OGLE-2003-BLG-84 is a free-floating
planet candidate.

	Event EROS-BLG-29 suffers from the ecliptic degeneracy.

\section{Likelihood Mass Analysis
\label{sec:likelihood}}

	Even with a precise measurement of $\bpi_\E$,
one cannot generally determine the lens mass $M$
unless $\theta_\E$ is also measured (see eqs.~[\ref{eqn:pie}]
and [\ref{eqn:tethetae}]).
With the possible exception of OGLE-2003-BLG-84,
none of the events analyzed in this
paper have $\theta_\E$ measurements.

	Instead we can estimate the likelihood of the
lens mass using a prior mass function together with prior
probability distributions for
the positions and velocities
of the source and lens.
Our model of the Milky Way consists of a double
exponential disk with a \citet{hangould96,hangould03} barred bulge.
We adopt a $R_\sun=8$ kpc for the Galactocentric distance of the Sun.
Our disk has a scale length of 3500 pc and a scale height of 325 pc.
We adopt a Solar velocity ${\bdv v}_\sun = (10,225,7)\,\kms$
in the Galactocentric, rotation, and north polar directions
and disk velocity dispersions of $(40, 30, 20)\,\kms$ in these
directions. We assume a mean disk rotation of 214 $\kms$
(220 $\kms$ rotation, less 6 $\kms$ asymmetric drift).
The bulge stars are assumed to have zero mean velocity
with dispersion of 80 $\kms$ in each direction.
By itself, this would not be a self-consistent model of the Galaxy. Note that,
\begin{equation}
G \Sigma h = k \sigma^2,
\label{eqn:surfacedensity}
\end{equation}
where $\Sigma$ is the surface density of
the disk, $h$ is the scale height, $\sigma$ is the velocity dispersion,
and $k$ is a constant of order unity.  Therefore, $h$ and $\sigma$ cannot
be constants.  We adjust these by making
\begin{equation}
\sigma \propto \Sigma^{1/3},\qquad h \propto \Sigma^{-1/3}.
\end{equation}
Since the bar most likely formed from a disk instability there should be
paucity of disk stars in the bar region.  Therefore, we removed the inner
disk from our model within 2 kpc of the Galactic center.  We find this has
little effect on the likelihood calculations.
Furthermore, we include radial velocity measurements \citep{cavallo02} to
constrain the
source position on events MACHO-98-BLG-6, MACHO-99-BLG-1, MACHO-99-BLG-8,
and MACHO-99-BLG-22.

The Sgr dwarf galaxy can contribute to microlensing events
and is included as a possible source. It is at a distance of
26.3 kpc \citep{monaco04}, and its stars are assumed to be moving
$2.2\,\masyr$ toward Galactic north \citep{ibata97} with negligible dispersion.
Sgr dwarf RR Lyrae stars are $\sim 2.6\%$ as numerous as bulge
RR Lyrae stars in the MACHO fields \citep{alcock97a}.
This factor is included in our model.

	We model $\Gamma_{\rm prior}$, the expected microlensing rate 
prior to making an observation,
\begin{equation}
\frac{d^5 \Gamma_{\rm prior}}{dM dx_l dx_s d^2\bmu }
 = \rho_{l}(x_l) x_l^2 2\theta_\E(M,x_l,x_s) \mu_\rel
   \phi(M)
   \rho_s(x_s) x_s^2
   f(\bmu;x_l,x_s)
   \biggl(\frac{x_s}{R_0}\biggr)^{-\beta}
\label{eqn:microrate}
\end{equation}
where $x_l$ and $x_s$ are the distances to the lens
and source respectively, $\rho_{l}(x_l)$ and 
$\rho_{s}(x_s)$ are the number density of each,
$f(\bmu;x_l,x_s)$ is the
distribution of relative proper motions, $\bmu$, given the
source and lens positions,
and $\phi(M) = dN/dM$ is the mass function (MF) of lenses
with normalization $\int \phi(M)dM = 1$.
Equation (\ref{eqn:microrate}) is simply a rate derived from
\begin{equation}
\Gamma = n \sigma v_T,\qquad
n = \rho_l(x_l) dx_l \phi(M) dM,\qquad
\sigma = 2 \theta_\E x_l,\qquad
v_T = \mu_\rel x_l,
\label{eqn:nsigv}
\end{equation}
where $n$ is the number density, $\sigma$ is the cross
section, and $v_T$ is the transverse velocity of the lens.
The additional factors $f(\bmu,x_l,x_s) d^2\bmu$ and
$\rho_{s}(x_s) x_s^2(x_s/R_0)^{-\beta}dx_s$ account for the
distributions of lens-source relative proper motions and
source distances.  Here, $\beta$ is a
parameter that characterizes the selection effects from
more distant, fainter source stars. We use $\beta = 1$ for
our calculations but find that changing it has little effect
on the results.  

	The result of the observations is to measure $\bvt$ and $\ret$,
which yields a trivariate error distribution function
$G(\ret-\tilde r_{\rm E,obs},\bvt-\tilde {\bf v}_{\rm obs})$ relative
to the best-fit values $\tilde r_{\rm E,obs}$ and $\tilde {\bf v}_{\rm obs}$.
The product of this function with equation (\ref{eqn:microrate})
gives the posterior rate $\Gamma_{\rm post}$, which we integrate
over all variables except $M$,
\begin{equation}
\frac{d\Gamma_{\rm post}}{dM}= \int
\frac{d^5 \Gamma_{\rm prior}}{dM dx_l dx_s d^2\bmu }
G(\ret-\tilde r_{\rm E,obs},\bvt-\tilde {\bf v}_{\rm obs})
dx_l dx_s d^2\bmu 
\label{eqn:massrate}
\end{equation}

Before performing the integration, we must switch integration variables
$\bmu\rightarrow \bvt$ and $x_l\rightarrow \ret$.  
We multiply by the Jacobian,
\begin{equation}
\frac{dx_l}{d\ret}\,{d^2 \bmu\over d^2\bvt}
= \frac{2x_l^2 \pirel}{\au \ret}\biggl({\pirel\over \au}\biggr)^2,
\label{eqn:jacob}
\end{equation}
which yields,
\begin{equation}
{d\Gamma_{\rm post}\over dM}
= 4\phi(M)\int dx_s d\ret d^2\bvt
\rho_l(x_l) x_l^4 {\biggl(\frac{\pirel}{\au}\biggr)}^5
           \tilde v \rho_s(x_s) x_s^2
           \biggl(\frac{x_s}{R_0}\biggr)^{-\beta}
           f(\bmu;x_l,x_s)
           G(\ret-\tilde r_{\rm E,obs},\bvt-\tilde {\bf v}_{\rm obs}),
\label{eqn:gamintermed}
\end{equation}
where $\pirel/\au = (x_l^{-1}-x_s^{-1})$, $\bmu=\bvt\pirel/\au$, and
$x_l$ itself
are implicit functions of $\ret=(\kappa M/\pirel)^{1/2}$,
$\bvt$, and $x_s$.
Finally, we assume that over the range that $G$ is substantially
different from zero, the remainder of the integrand varies
relatively little.  It is then appropriate to replace $G$
by a 3-dimensional $\delta$-function.  Integration of equation
(\ref{eqn:gamintermed}) then yields
\begin{equation}
{d\Gamma_{\rm post}\over dM}
= 4\phi(M)\int dx_s 
\rho_l(x_l) x_l^4 {\biggl(\frac{\pirel}{\au}\biggr)}^5
           \tilde v_{\rm obs} \rho_s(x_s) x_s^2
           \biggl(\frac{x_s}{R_0}\biggr)^{-\beta}
           f(\bvt_{\rm obs}\pirel/\au;x_l,x_s),
\label{eqn:gamintfinal}
\end{equation}
where again, $x_l$  and $\pirel$ are implicit functions of
$\ret$, $x_s$ and $M$.

\subsection{Mass Function}

	The MF of the bulge main sequence (MS) has been
measured in both the optical
\citep{holtzman} and the infrared \citep{zoccali} using
{\it Hubble Space Telescope (HST)} observations.  For purposes of this paper,
we adopt a MS MF that is consistent with those measurements
(but without corrections for binaries),
\begin{equation}
{d N\over d M} = k\biggl({M\over \mbr}\biggr)^\alpha,\qquad
\mbr = 0.7\,M_\sun,
\label{eqn:msmf}
\end{equation}
where $k$ is a constant, and
\begin{equation}
\alpha = -1.3\quad (0.03\,M_\sun<M<\mbr),\qquad
\alpha = -2.0\quad (\mbr<M\la M_\sun).
\label{eqn:powerlaws}
\end{equation}
The upper of limit of $\sim 1\,M_\sun$ is the approximate position of the 
turnoff.  
The lower limit is arbitrary and simply extends the slope of the 
\citet{zoccali} observations from their last measured point at
$0.15\,M_\sun$ into the brown dwarf regime.
The MF may well extend even further,
but events showing significant parallax distortion are typically too long
to be caused by lower mass lenses.  

	We assume that all MS stars in the range $1\,M_\sun<M<8\,M_\sun$
have now become WDs, and that the total number can be found by extending
the upper MS power law $\alpha=-2.0$ through this higher-mass regime.
That is, $N_{\rm WD}= (7/8) k \mbr^2/M_\sun$.  Of course, there is no evidence
whatever that the slope does continue in this regime.  A more popular slope
is the Salpeter value $\alpha=-2.35$.  Had we chosen this steeper slope, 
the estimate
for $N_{\rm WD}$ would be reduced by a factor 0.80.  For the distribution
of WD masses, we adopt the MF shown in Figure 11c of \citet{bragaglia}
based on observations of 164 hot WDs.  We assume that all MS stars 
$8\,M_\sun<M<40\,M_\sun$ become NSs, with masses that are centered
at $M=1.35\,M_\sun$ and with Gaussian dispersion of $0.04\,M_\sun$
\citep{thorsett}. We assume that all MS stars 
$40\,M_\sun<M<100\,M_\sun$ become BHs, with masses that are centered
at $M=7\,M_\sun$ and with Gaussian dispersion of $1\,M_\sun$.
We also assume that the power law $\alpha=-2$ extends
throughout this entire regime.

The lens contributes to the baseline light of an event.
Thus the blending of the event gives an upper limit on the brightness of
a lens.  This brightness limit provides an upper limit on the mass of a main
sequence lens.
We model extinction as a double exponential with a dust scale length of 
3500 pc and a scale height of 130 pc with a local extinction of 0.4 mag/kpc.
At each source and lens distance we use the
\citet{cox99} mass-luminosity relationship to cut off the MS portion of the
MF.

\subsection{Likelihood Results}

The relative $\Gamma$ likelihoods for
each type of lens are shown in Figures \ref{fig:likeli-mb99-22}-
\ref{fig:likeli-mb98-6}, where they are normalized
to unity and so expressed as a probability distribution function. The
dotted line is the expected distribution of each event from out Galactic
model.
 
Figure \ref{fig:combinedgamma} shows a composite of the expected microlensing
rate based on our Galactic model for the non-(BH/BD/xallarap) events.
This figure
shows how the events in this paper are not typical, but instead have
lenses that are relatively more massive and closer to us.
Also, tables in the electronic version show the contribution of the
various lens and source
populations to the total likelihood. ``Both'' refers to both disk
and bulge lenses, and ``All' refers to disk, bulge, and Sgr sources.
The displayed numbers for each solution are weighted by $\exp(-\dchi/2)$ such
that the totals for each solution all add to make up the combined table.
The weight factor is listed next to the solution number.

\subsection{Black Hole Candidates
\label{sec:bh}}

	Three parallax events have been identified in the literature
as black-hole candidates, MACHO-96-BLG-5, MACHO-98-BLG-6, and
MACHO-99-BLG-22.  We assess these candidates by evaluating the
relative probability that they lie in one of four stellar
classes: main-sequence stars and brown dwarfs (MS), white dwarfs (WD),
neutron stars (NS), and black-holes (BH).  When there is more than
one solution, we evaluate each solution separately and then combine
them, weighting each by $\exp(-\dchi/2)$.
For each solution we place an upper limit on the mass of a MS lens
based on blending of the light curve.  There are additional photometric
constraints available for MACHO-96-BLG-5.

	\citet{bennett02} obtained {\it HST} images of MACHO-96-BLG-5
to constrain the blending of the source star.
They found that 31\% of the total flux was from the source star
or from other stars (such as the lens) that are very closely
aligned with it.
However, as they note, this does not rule out that another blended star
(possibly the lens) is unresolved by HST.
The best fit parallax solution yields a blending of 87\%,
implying that the source itself comprises 13\% of the baseline light.
So for this event, we place an upper constraint on a main sequence mass
based on the lens contributing no more than 18\% (=31\%-13\%) of the baseline
light instead of the 87\%
found by just fitting the parallax model.
We weight each solution by $\exp(-\dchi/2)$ and find
relative likelihoods of MS:WD:NS:BH :: 31:19:14:37.
Nearly all the BH probability is from disk lenses and bulge sources.
This combination accounts for only about 37\% of the total likelihood.

	MACHO-98-BLG-6 has two solutions, but the second has
little weight, $\exp(-\dchi/2)=3\%$.  We find
$\dchi$-weighted
relative likelihoods of MS:WD:NS:BH :: 58:26:13:2.
Like, the previous event, the BH probability is from disk lenses and
bulge sources. This combination of source and lens locations
makes up only 21\% of the likelihood for this event.

	MACHO-99-BLG-22 is a strong BH candidate with
relative likelihoods of MS:WD:NS:BH :: 11:4:7:78.
Unlike the previous two events, the BH probability is dominated by
bulge-bulge lensing. This is due to its unusually large $\ret=30\au$, and
high $\tilde v=83\,\kms$.

	Of course these likelihood ratios depend on the priors
embedded in the assumed mass function.  In particular, if the
fraction of black holes in this mass function were increased, the
black-hole likelihood would rise correspondingly.  However, if
the prior on the black-hole fraction were altered within a
plausible range, it would not materially affect the interpretation of
MACHO-98-BLG-6.  That is, even if the
assumed density of black-holes were doubled, this would only
increase the black-hole likelihood of MACHO-98-BLG-6
from 2\% to 4\%.  However, if this density were cut in half, 
this would decrease the black-hole likelihood of MACHO-99-BLG-22
from 78\% to 64\% and of MACHO-96-BLG-5 from 37\% to 22\%.

\citet{smith05} examine these three events and find MACHO-99-BLG-22 and
MACHO-96-BLG-5 are inconsistent with their simulation which does not
include stellar remnants (see their Fig. 12).
Event MACHO-98-BLG-6 is consistent with their
simulation.  Their results are compatible with our likelihood analysis.

\subsection{Free-Floating Planet Candidate
\label{sec:planet}}

	OGLE-2003-BLG-84 reaches a peak magnification $A_{\rm max}\sim
10$.
We therefore incorporate finite-source effects into the fit and find
an improvement $\Delta\chi^2=8.81$ for
$\rho\equiv\theta_*/\theta_{\rm E} = 0.107\pm0.030$,
where $\theta_*$ is the angular-size of the source.  The source star
has a baseline magnitude $I=20.3$ with essentially no blending.
Although we do not have a color-magnitude diagram of the field,
this apparent magnitude is generally consistent with an upper-main
sequence star in the Galactic bulge.  Hence, in fitting for finite
source effects, we use a solar linear limb-darkening coefficient
(on the \citealt{albrow} system) $\Gamma_I=0.3677$, and we estimate
$\theta_* = r_\odot/R_0 = 0.58\pm 0.07\,\mu\rm as$.  This implies
$\theta_{\rm E} = \theta_*/\rho = 5.4\pm 1.7,\mu\rm as$.  The best-fit
parallax
when finite-source effects are included is hardly changed,
$\pi_{\rm E}=0.65\pm 0.15$, or $\tilde r_{\rm E} = 1.54\pm 0.39\,$AU.
Combining the measurements of $\theta_{\rm E}$
and $\pi_{\rm E}$ yields,
\begin{equation}
M = {\tilde r_{\rm E}\theta_{\rm E}\over \kappa} =
1.0\pm 0.4 \times 10^{-3}\,M_\odot,\qquad
\pi_{\rm rel} = \pi_{\rm E}\theta_{\rm E} = 3.5\pm 1.3 \mu\rm as.
\label{eqn:mpirel}
\end{equation}

	The implied Jupiter-like mass would be very exciting if
true.  However, the fact that this is only a $3\sigma$ detection
implies that caution is warranted.
The conjointly derived relative parallax measurement is
extremely small, implying a lens-source separation of only
$D_{\rm LS} = 224\,{\rm pc}(D_s/R_0)$, where $D_s$ is the distance
to the source and $R_0=8\,$kpc is the Galactocentric distance.
Such a small source-lens separation is a priori unlikely, but
cannot be strongly argued against on those grounds.  This is
because almost the only events that can give rise to both significant
parallax and finite-source effects are those with very small
$\theta_{\rm E}$ and fairly small $\tilde r_{\rm E}$.  These criteria
already imply very small (i.e., planetary) masses and generally
small $\pi_{\rm rel}$.  Hence, while the characteristics of this
event are intrinsically unlikely for an event chosen at random, they are
``normal'' for an event with measured $\theta_{\rm E}$ and $\tilde r_{\rm
E}$.

	The xallarap analysis ($\dchix=3.88$) of the event strongly
confirms that parallax is the dominant contributor to the light curve
distortions.  Under the assumption that the lens is indeed a planet,
this can be used to place a rough lower limit on the
planet's orbital distance $a_{\rm planet}$ from a possible host star.
We assume the light-curve distortion from the planet's acceleration is
less than $1/4$ of the contribution from the Earth's motion; otherwise
the xallarap
solution would be pushed away from the Earth parameters. However, since
$D_S/D_L \sim 8\,{\rm kpc}/224\,{\rm pc} \sim 36$, the effect of the
planet's acceleration on the light curve is effectively multiplied 36
times. Hence, the acceleration of the planet must be $4\times36=144$
times smaller than that of the Earth. This implies a semi-major axis
$a_{\rm planet} > 12\,\au (M_*/M_\sun)^{1/2}$, where $M_*$ is
the mass of the host star.

	While there is some evidence that this event
is caused by a free-floating planet or by a planet at very wide separation
from its parent star, there is no clear way
to confirm this conclusion.  First, the inference that there are
finite-source effects rests on a relatively low $\dchi$.  Second,
there are no independent data sets that could confirm this measurement.
Finally, there is no way to independently verify that there is a
free-floating planet at a distance $D_l\sim 8\,$kpc.  On the other hand,
if the planet is bound to a star at wide (but not too wide) separation,
then the star itself may eventually give rise to a second bump in the
light curve.  Unfortunately, the characteristics of this bump cannot
be accurately predicted.  If the star is, say, $0.3\,M_\odot$, then the
stellar Einstein radius would be about 17 times larger than that
of the planet, so one might then expect the heliocentric timescale
to be 17 times longer than the $t_{\rm E}\sim 430$ days found for the
lens.  However, if the planet were orbiting at $5\,\rm km\,s^{-1}$,
this would produce a difference in projected velocities of about
$(8000/226)\times 5\,\kms=180\,\kms$
between the planet and star.  Since the projected velocity of the lens
is only $6\,\rm km\,s^{-1}$, this would reduce the predicted stellar
Einstein
crossing time by a factor of $180/6=30$ times from $17\times 430=7310$
days
to 244 days.  Thus, the stellar bump could occur a few years (i.e., a
few Einstein timescales) after the planetary bump and be of similar
duration to it.  In any event, if this second bump does occur, it
will confirm the planetary nature of the event. But absent of that,
this ``planet'' must remain a candidate.

\subsection{Brown Dwarf Candidate MACHO-179-A}
\label{sec:BDcandidate}

The brown dwarf candidacy of MACHO-179-A rests primarily on its
small projected Einstein radius, $\tilde r_{\rm E} = 0.23\pm 0.05\,\rm
AU$.
 From equation (\ref{eqn:pie}), this implies a lens mass
\begin{equation}
M ={\pi_{\rm rel}\over \kappa \pi_{\rm E}^2}\rightarrow
0.065\,M_\odot\,{100\,{\rm pc}\over D_l},
\label{eqn:mbd}
\end{equation}
with an error of about 0.17 dex.  In the last step, we have made the
approximation that $D_l\ll D_s$, which is certainly true unless
the source lies well in the foreground ($D_s\la 500\,{\rm pc}$)
or the lens is of extremely
low mass $(M\la 10^{-3}\,M_\odot)$.  Then, from the projected
velocity $\tilde v = 26.6\,\rm km\,s^{-1}$, the proper motion is
\begin{equation}
\mu = \tilde v{\pi_{\rm rel}\over \rm AU}
\rightarrow 56{\rm mas\over yr}\,{100\,{\rm pc}\over D_l}
= 56{\rm mas\over yr}\,{M\over 0.065\,M_\odot}.
\label{eqn:mubd}
\end{equation}
Hence, if the lens is a star ($M>0.08\,M_\odot$) rather than a brown
dwarf, it is quite close ($D_l\la 80\,\rm pc$) and has an extremely
high proper motion, $\mu\ga 69\,\rm mas\,yr^{-1}$.  Such a star
should be quite easy to spot.  For example, even extreme red dwarfs
($M\ga 0.08\,M_\odot$) have $M_I\la 14$, and so $I\la 18.5$.  An old
white dwarf of $M=0.6\,M_\odot$ and $M_I\sim 15$, would be at 10 pc,
and so quite bright, $I\sim 15$.

There are already significant constraints on any such lenses
from the light curve.  The baseline flux is $R=19.6$, and the
blending fraction is $\eta_b = 0.32\pm 0.23$.  Hence, the
lens brightness is limited by $R_l \ga 20$.  This clearly rules out
white dwarfs and permits only the most extreme M dwarfs, with
$M\la 0.1\,M_\odot$.  Even these could easily be found by direct
CCD imaging of the field.  The very high predicted proper motion
of the lens guarantees that it could be unambiguously identified
if it is luminous.

If this option is excluded by future observations, then there remain
only four possibilities: the lens is a brown dwarf, a neutron star, or
a black-hole, or the light-curve distortion is not due to parallax
at all, but to some other effect, most likely xallarap.  We examine
the three alternatives to a brown dwarf in turn.

A black hole would lie at only 1 pc. It would have an angular Einstein
radius $\theta_{\rm E}\sim 0.\hskip-2pt ''25$ and a proper motion of
about $6''\,\rm yr^{-1}$.  It should therefore have given rise
to a substantial number of microlensing events both during the
original MACHO observations and also during observations of this
field by OGLE-III.

A neutron star would lie 5 pc and have
$\theta_{\rm E}\sim 50\,\rm mas$.  While fairly large, this would not
be large enough to guarantee other microlensing events during the
periods of observations.  However, other effects might be observable,
such as X-ray emission due to interactions of the NS with the interstellar
medium.

Xallarap can easily produce distortions that are misinterpreted as
due to parallax with small $\tilde r_{\rm E}$.  Indeed, the
event OGLE-2002-BLG-100 also has a fairly small
$\tilde r_{\rm E}\sim 0.46\,\rm AU$, which, as we show in
\S~\ref{sec:xallarap},
is misinterpreted xallarap.   By contrast, the
xallarap analysis for MACHO-179-A shows that the xallarap parameters
closely mimic the Earth's orbital parameters, indicating that it is
unlikely, but not impossible, that the light-curve distortions are
due to xallarap.

Moreover, the xallarap interpretation would imply other, possibly
observable effects.
Because these parameters indicate a source-binary orbital period of
1 year and a nearly edge-on inclination, the source should have
an annual velocity semi-amplitude of
$v\sim 30\,{\rm km\,s^{-1}}q(1+q)^{-2/3}$, where $q$ is the ratio
of the source companion mass to the total mass of the binary.  Thus,
unless the source companion is extremely light, it should be possible
to test the xallarap hypothesis directly from RV measurements, although
the faintness of the source would make this quite difficult.

Finally, we note that if the lens is a brown dwarf, it may also be
possible
to directly image it in the infrared.  This will of course depend on
its luminosity, which falls rapidly with both increasing age and
decreasing
mass.

\section{Xallarap Analysis
\label{sec:xallarap}}

The parallax effect is not the only way that
a microlensing light curve can be distorted.
The source may be part of a binary in which the
acceleration of the source is causing the observed light-curve
asymmetry \citep{griesthu,hangould97,pac97}.
This effect is often called ``xallarap''.
For each of these events, we search the class
of xallarap solutions in which the source is in a circular orbit
and its companion is either not lensed or too faint to contribute to the 
light curve.
We search periods from 215 to 515 days (in steps of 20 days),
every possible phase, and every orbital inclination (both in
steps of 5 degrees). Additional
periods are searched when the best fit
period is near the lower or upper limit of the range probed.

	If the light-curve distortion is actually due to parallax
rather than xallarap, we expect the best-fit xallarap solution
to closely mimic the parallax solution.  That is, we expect
the best-fit period to be the period of the Earth (1 year), 
the best fit phase to be the ecliptic longitude of the event
(typically $\lambda=270^\circ$ for bulge events), and the best fit inclination 
to be the (complement of) the ecliptic latitude of the event (typically 
$-11^\circ\la \beta_{\rm ec}\la -3^\circ$ for bulge events).
There exists a perfect north-south latitude degeneracy
when the sign of $u_0$ and
$\pi_{\E,N}$ are both changed simultaneously. Therefore we use
the degenerate solution with the opposite $u_0$ as a seed for the northern
latitudes. We search phases corresponding to ecliptic longitudes
$180 \le l_{\rm ec} \le 360$ since there is an exact degeneracy in the
supplementary angles.
Note that the expected improvement
for allowing the three extra xallarap parameters is small,
$\langle\dchi\rangle=3$.

	For four of the events, we find $\dchi<2$ (see Table 1),
thus clearly confirming the parallax interpretation.
Conversely, three of the events have $\dchi>27$ which is
a strong indication that these light curves have been distorted by
xallarap. Another seven events have
$ 3 < \dchi < 9 $. The remaining eight events have
$ 10 < \dchi < 25 $. Since only three extra degrees of
freedom are introduced in the xallarap analysis, it is
surprising to find so many large improvements in $\chi^2$.
The majority of these may be the result of small systematic effects
in the data that are more closely modeled by invoking these extra parameters.
One possible such ``systematic effect'' is a minor xallarap perturbation
on a distortion that is predominately caused by parallax.  That is, if
the source of a parallax event were a member of a binary, the small
resulting xallarap effect could perturb the parallax solution, pushing
it slightly away from Earth-like parameters.

	One way to examine the possible influence of xallarap is to consider
how reasonable the parameters are from the parallax solution versus the
xallarap solution. The Einstein ring size projected into the source
plane $\rehat$ is related to $\ret$, the Einstein ring size in the
observer plane inferred from the ``parallax interpretation'' of the event, by
\begin{equation}
\hat r_{\rm E} = {q\over (1+q)^{2/3}}
\biggl[{M_s\over M_\sun}\,
\biggr({P\over {\rm yr}}\biggr)^2\biggr]^{1/3}\tilde r_{\rm E},
\label{eqn:rehat}
\end{equation}
where $q$ is the mass ratio of the source binary, $M_s$ is the mass of the
source being lensed, and $P$ is the source period.
Note that the factor in front has a maximum plausible value of
$2^{-2/3}=0.63$, under the reasonable assumption that $q \le 1$.
Another way to check for xallarap is to obtain radial
velocity measurements of the source to confirm or discount its binary
nature. We now discuss individual events with $\dchi>4$.

	MACHO-96-BLG-5 has $\dchix=12.68$, but the best solution has
a period of 365 days with a phase and inclination very similar to
to Earth's parameters (i.e. parallax).
This is because the xallarap improvement is quite small compared to the
parallax improvement $\dchip=2107.86$.
That is, xallarap effects are at most a minor perturbation on parallax,
which predominates overwhelmingly.

	MACHO-98-BLG-6 has $\dchix=15.89$. The best solutions
tend to clump around periods of 400 days, but there does exist a solution with
$\dchix=13.22$, whose period (365 days) and complement of 
inclination ($-5^\circ$) both agree with the parallax solution and whose
phase differs from the parallax solution by only $20^\circ$. 
Again $\dchix \ll \dchip=
557.71$, meaning that xallarap is at most a minor perturbation.

	MACHO-99-BLG-1 has $\dchix=15.58$. 
The phase and inclination parameters have a lot of
scatter among the best xallarap solutions, but they are clumped around
the Earth's parameters. The best periods are near 750 days.
However, the parallax model improves the upon the standard model
considerably ($\dchip=1639.02$). The best xallarap solution improves
only a small fraction better than the parallax model.
Therefore we conclude this event is not strongly affected by xallarap.

	MACHO-99-BLG-8 has $\dchix=14.72$ with parameters
similar to Earth's. 
Again the $\dchix\ll\dchip=1190.30$.
Therefore we conclude this event is not strongly affected by xallarap.

	MACHO-179-A has a relatively small $\dchi=8.76$ 
and its best-fitting solutions clump near the parallax solution.
We tentatively conclude that this event is not strongly affected by xallarap.
However, in view of the striking implications of the parallax interpretation,
an additional investigation is warranted (see \S~\ref{sec:BDcandidate}).

	MACHO-98-BLG-1 with $\dchix=112.24$ is the strongest
xallarap candidate of the events in this paper. All three parameters
differ significantly from the parallax solution with best-fitting
periods near 425 days.

	MACHO-95-BLG-27 has $\dchix=17.38$ with xallarap parameters
inconsistent with Earth's. The xallarap solutions strongly favor periods
near 410 days and inclinations $\sim45\arcdeg$.
Also the xallarap $\chi^2$ improvement is significant relative to the
parallax improvement over the standard model ($\dchip=190.69$).
We conclude this event is strongly
affected by xallarap.

	MACHO-99-BLG-22 has $\dchix=4.28$. While the xallarap
solutions do have periods near 365 days, the inclination parameters tend
toward face on solutions. The phase is not determined well from the fits.
Again the $\dchix\ll\dchip=640.30$.
We conclude this event is not strongly affected by xallarap.

	OGLE-1999-BLG-19 has $\dchix=19.75$, but all the best
solutions have parameters very close to the parallax parameters.
\citet{smith02} also found that their xallarap model mirrored the Earth's
orbital parameters.
This is another case for which $\dchix \ll \dchip=10506$, so
this event is not strongly affected by xallarap.

	OGLE-1999-CAR-01 has $\dchix=5.13$ with parameters consistent
with parallax. 
Therefore we conclude this event is not strongly affected by xallarap.

	OGLE-2000-BLG-43 has $\dchix=6.13$ and has orbital
parameters consistent with Earth's, while
$\dchix\ll\dchip=3519.65$.
Therefore, this event is not strongly
affected by xallarap.

	The OGLE event sc33\_4505 has $\dchix=29.17$ and
is a strong xallarap candidate. The best solutions have periods near
200 days. The xallarap solution clearly
fits the data better as seen in Figure \ref{fig:sc33lc}.

	OGLE-2002-BLG-100 has $\dchix=27.62$, which is substantial
compared to $\dchip=139.38$.
This event is likely
affected by xallarap. The best solution has a period of 125 days, although the
phase and inclination are consistent with those of the Earth.
If the parallax solution were accepted at face value, its
$\pi_{\rm E}\sim 3$ would imply that the lens was a brown-dwarf
candidate, just as with MACHO-179-A.  This has low prior probability,
but is not implausible.  We should then also consider how plausible
is the xallarap solution.  First, we note that the parallax parameter
in the best xallarap solution is $\pi_{\rm E}\sim 0.3$ (compared
to 3 in the parallax solution).  Therefore, according to
equation (\ref{eqn:rehat}), the size of the Einstein ring
projected on the source plane is
$\hat r_{\rm E}\sim 1.6 q(1+q)^{-2/3}\,\rm AU$, which ranges from
0.15 AU to 1.0 AU for $0.1<q<1$.  Note that $\hat r_{\rm E}=1\,\rm AU$
implies $D_{LS} \simeq 122\,{\rm pc} (M_\odot/M)$, where
$D_{LS}= D_S - D_S$ and where we have assumed $D_{LS}\ll D_S$.
These masses and separations are fairly representative for bulge
populations.

	OGLE-2002-BLG-334 has $\dchix=5.98$ and favors periods
near 425 days and phases and inclinations inconsistent with parallax.
The small $\dchix$ leads us to conclude it is not strongly affected
by xallarap, especially since $\dchip=612.88$.
The differences must be due to systematic effects better
modeled by the extra three parameters.

	OGLE-2002-BLG-61 has $\dchix=11.66$ and favors periods
near 320 days with phases and inclinations inconsistent with parallax.
This $\chi^2$ improvement is significant relative to the parallax model's
improvement over the standard model ($\dchi=138.85$).
Also, the best xallarap solutions have $\sim0$ blending
which is more expected on this moderately bright event ($I_{\rm baseline}=17.4$).
We conclude that this event is affected by xallarap.

	OGLE-2003-BLG-188 has $\dchix=8.21$ and favors periods
near 440 days and inclinations inconsistent with parallax. However, the
$\dchi$ improvement is not very significant, and we have no clear reason to
conclude it is strongly affected by xallarap.

	OGLE-2003-BLG-32 has $\dchix=12.84$ and has orbital
parameters consistent with Earth's.
The improvement is minuscule compared to $\dchip=8042.12$,
and so it is not strongly affected by xallarap.

\section{Discussion and Conclusions
\label{sec:discussion}}

We have systematically studied 22 microlensing events with detectable
parallax ($\dchi>100$).
We outlined a procedure that easily identifies the three 
discrete degeneracies known to affect microlensing parallax events.
This procedure works provided the geocentric reference frame is used.
Surprisingly, we find 44 degenerate solutions among the 22 events in this
paper. These degenerate solutions need to be considered even in events
that appear to have a strong parallax detection.
For example, event MACHO-104-C has a $\dchi=1647.18$ parallax improvement
over the standard microlensing model and we still find 4 highly degenerate
solutions.
Six of our events
have $\dchi>1000$ parallax improvements over the standard microlensing model.
Only four of these have unique non-degenerate solutions (one of which is
strongly affected by xallarap rather than parallax).
We find no correlation between the number of degenerate solutions and
either the ecliptic latitude or $\dchip$.

We have reanalyzed three events previously identified as BH candidates
and taken into account the newly identified solutions.
We find the lens of event
MACHO-99-BLG-22 is a strong BH candidate, event MACHO-96-BLG-5
is a marginal BH candidate, and event MACHO-98-BLG-6 is a very weak BH
candidate.

\begin{figure}
\plotone{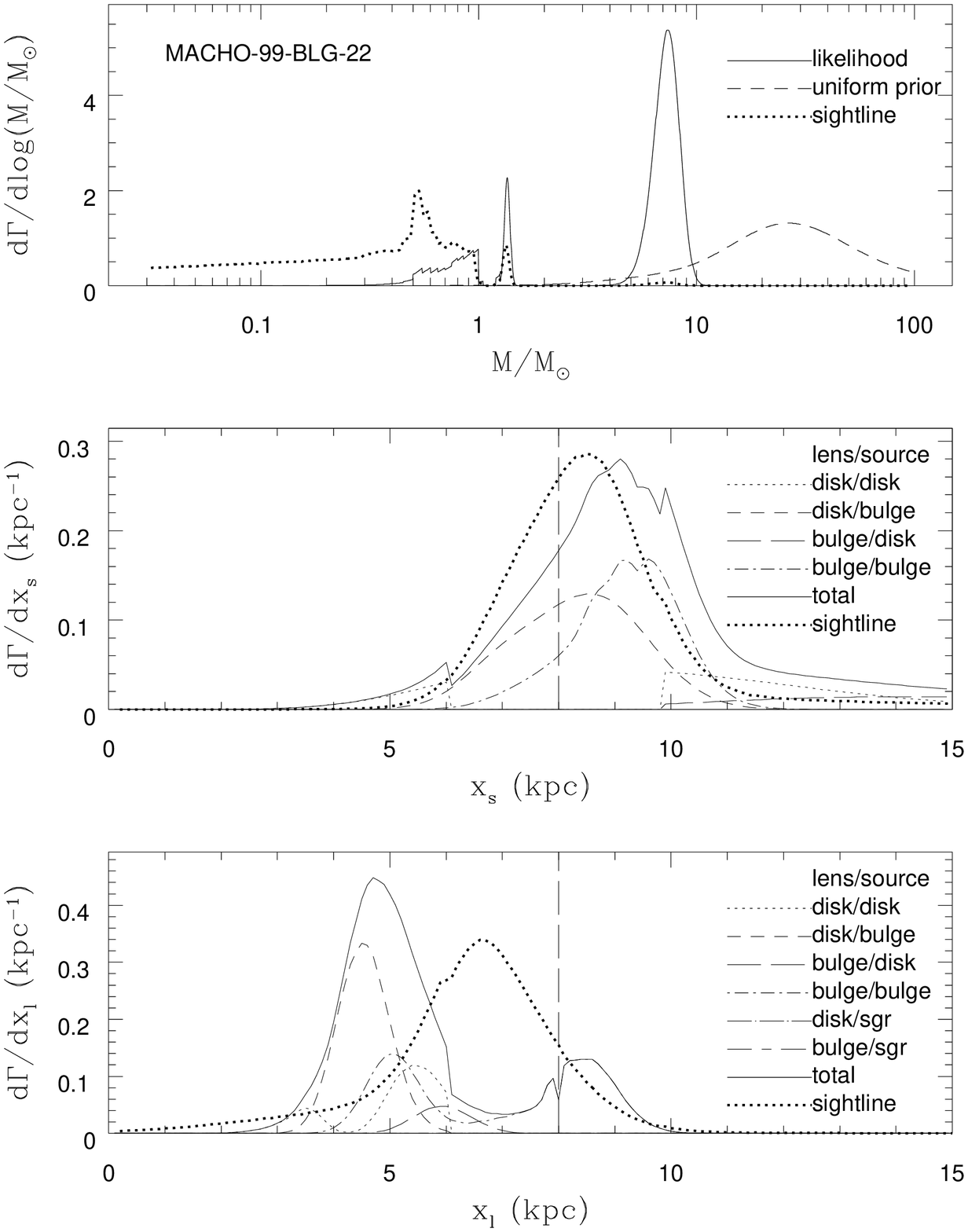}
\caption{Relative likelihoods versus mass, lens distance, and source distance
for MACHO-99-BLG-22.
In the upper panel, the solid curve is the result of our likelihood mass
calculations. The bold dotted line is the expected distribution along this
sightline from our
Galactic model. The dashed line is the mass likelihood with a uniform prior in
logarithmic mass. The middle and bottom panel give the likelihood of the
source and lens distances, respectively.
\label{fig:likeli-mb99-22}}
\end{figure}

\begin{figure}
\plotone{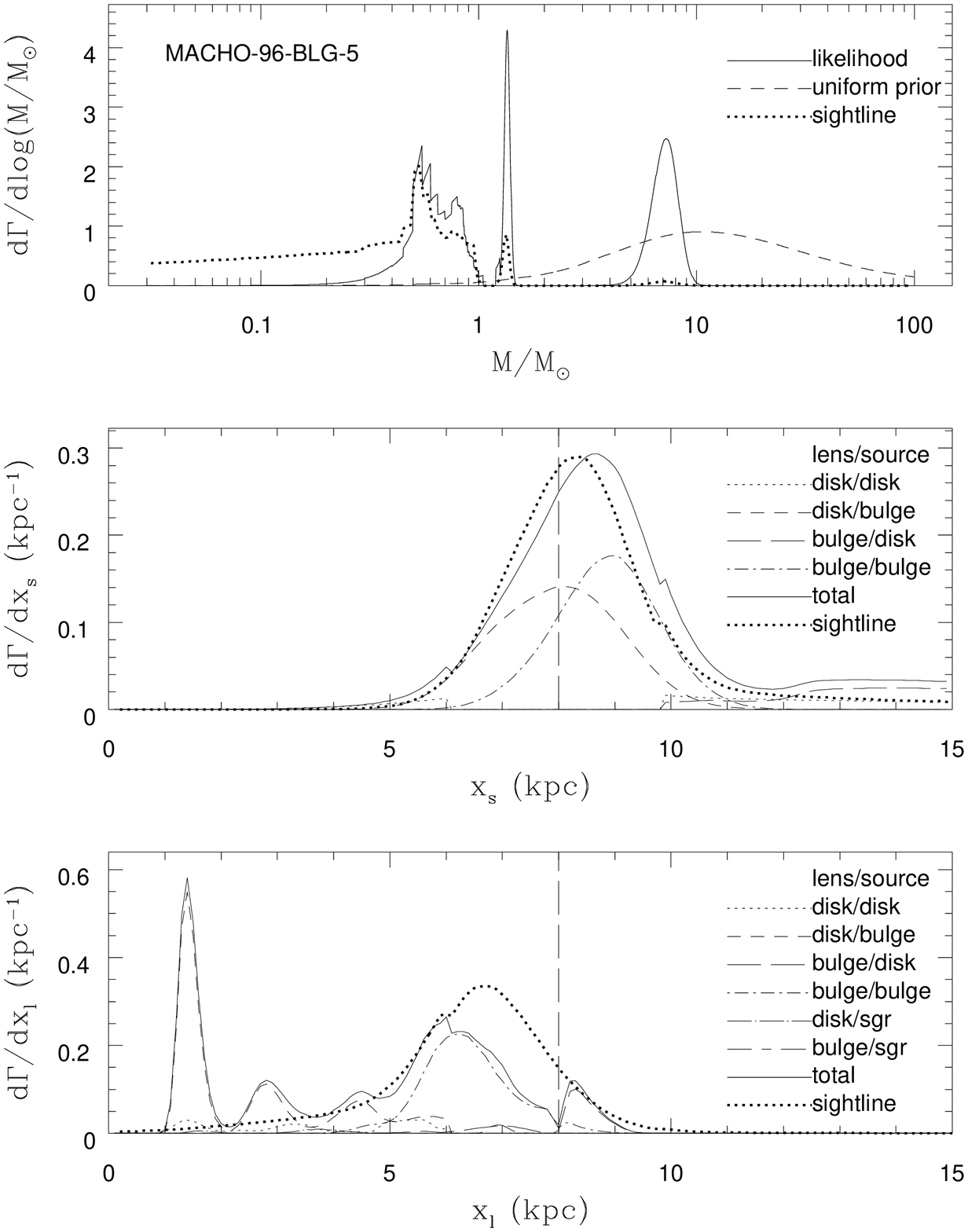}
\caption{Relative likelihoods versus mass, lens distance, and source distance
for MACHO-96-BLG-5.
In the upper panel, the solid curve is the result of our likelihood mass
calculations. The bold dotted line is the expected distribution along this
sightline from our
model. The dashed line is the mass likelihood with a uniform prior in
logarithmic mass. The middle and bottom panel give the likelihood of the
source and lens distances, respectively.
\label{fig:likeli-mb96-5}}
\end{figure}

\begin{figure}
\plotone{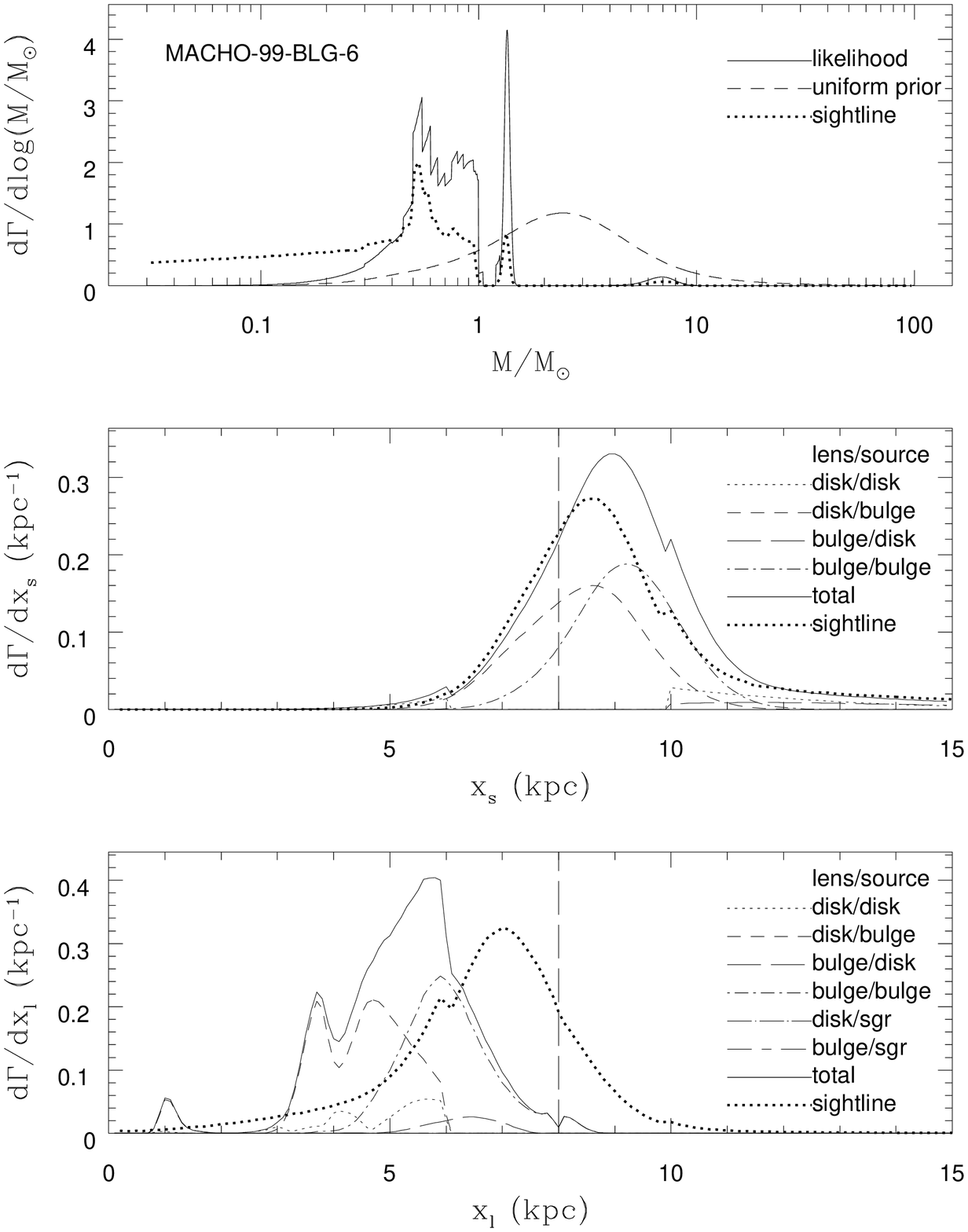}
\caption{Relative likelihoods versus mass, lens distance, and source distance
for MACHO-98-BLG-6.
In the upper panel, the solid curve is the result of our likelihood mass
calculations. The bold dotted line is the expected distribution along this
sightline from our
model. The dashed line is the mass likelihood with a uniform prior in
logarithmic mass. The middle and bottom panel give the likelihood of the
source and lens distances, respectively.
\label{fig:likeli-mb98-6}}
\end{figure}

\begin{figure}
\plotone{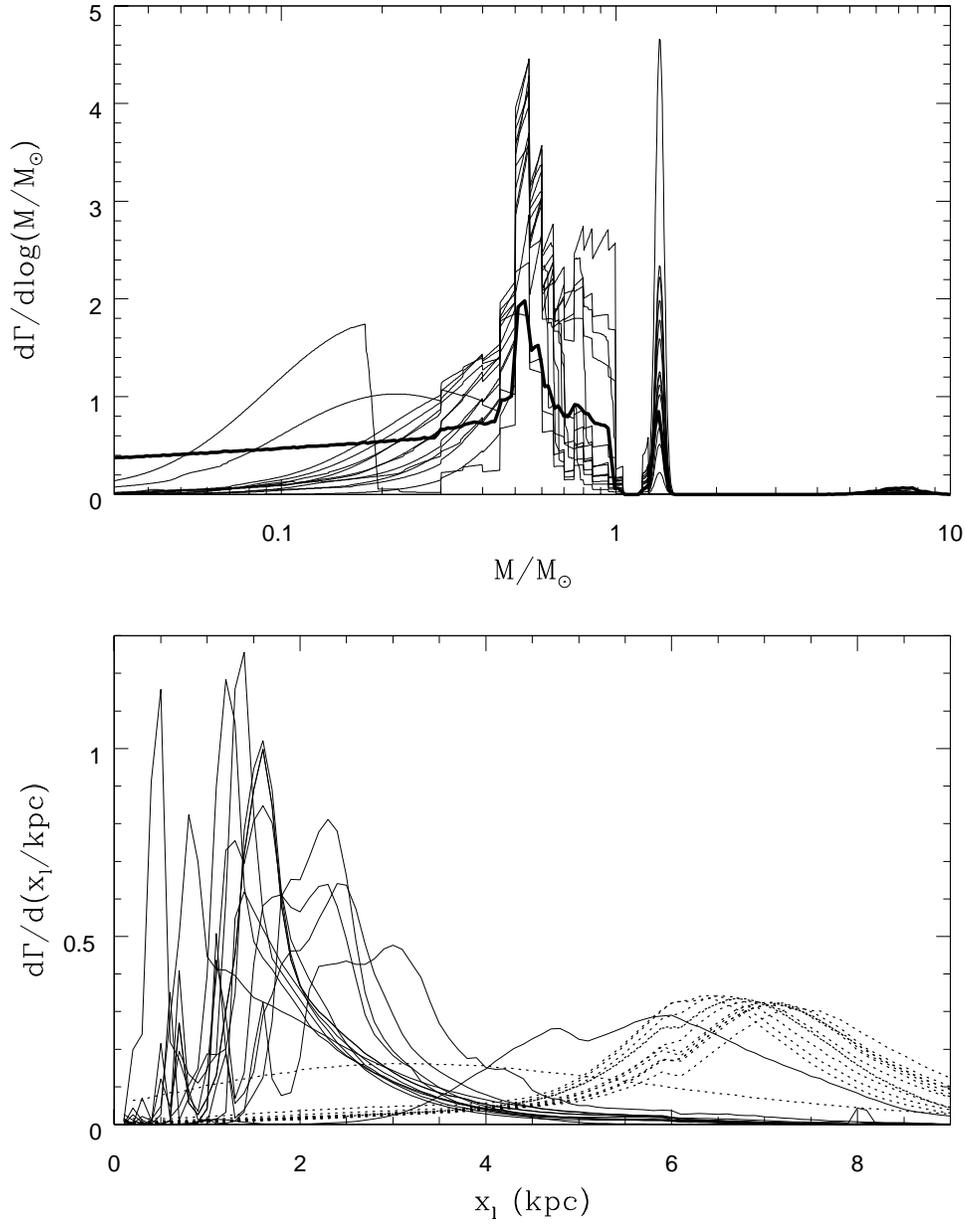}
\caption{The relative likelihood of all
12 non-(black-hole/brown-dwarf/planet/xallarap) events from this paper.
The bold curve in the upper panel is our adopted mass-function prior.
The bottom panel also shows the relative likelihoods for the same events,
but versus lens distance. The dotted lines are the expected distribution
of events
from our Galactic model in the same directions as these observed events.
These composite plots
illustrate how these parallax events have lenses that are closer and more
massive than average.
\label{fig:combinedgamma}}
\end{figure}

\begin{figure}
\plotone{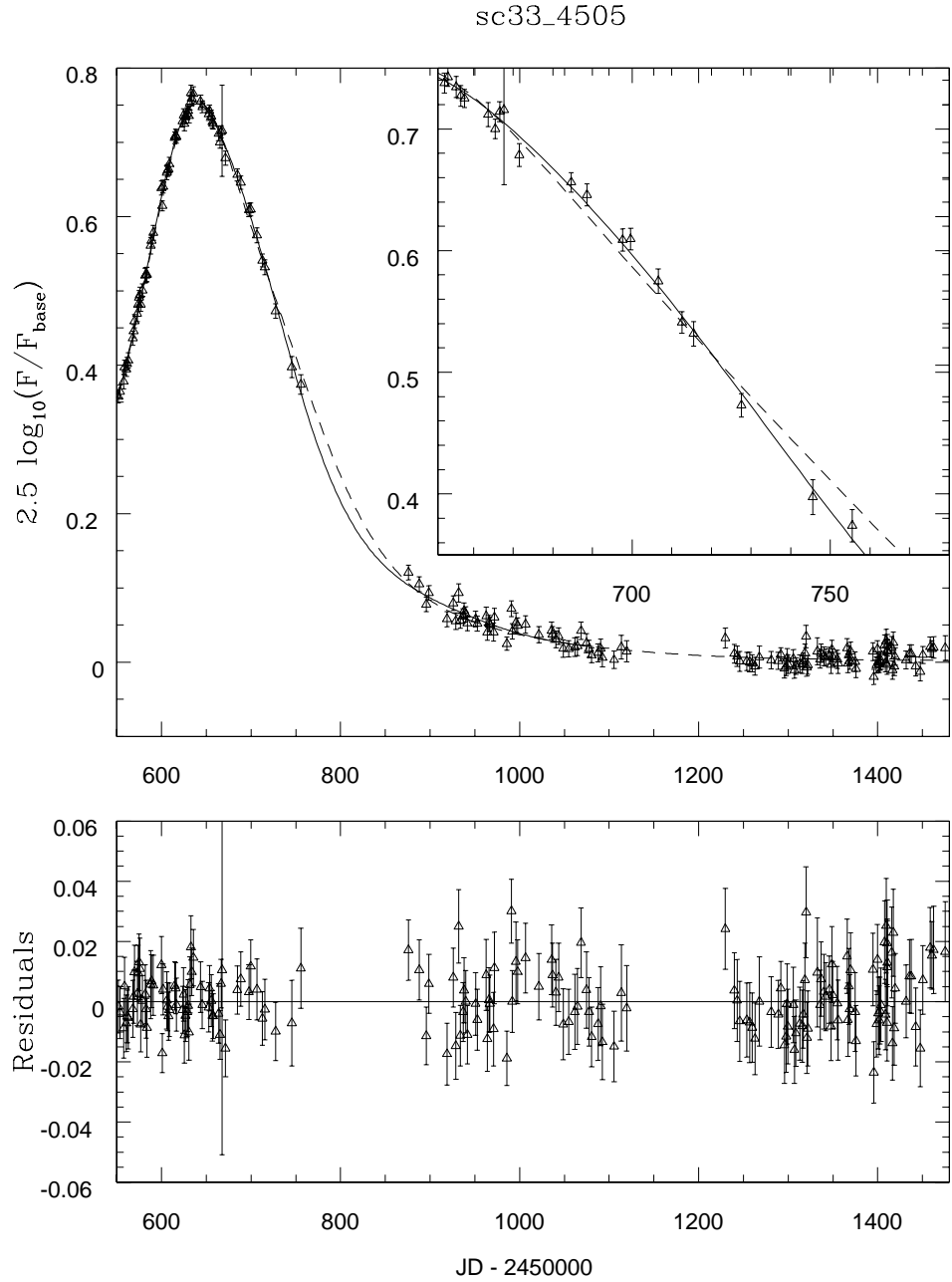}
\caption{OGLE-II light curve of sc33\_4505. The solid curve is the best fit
xallarap model with $\dchix=29.18$, and the dashed curve is the
best-fit parallax model. The upper right panel is an expanded view of the
light curve that shows how the data better fit the xallarap model.
The bottom panel shows the residuals of the xallarap model.
\label{fig:sc33lc}}
\end{figure}

	The adopted BH mass function ($M\sim7\,M_\sun$) is somewhat
arbitrary and is based entirely on BHs found in binaries.
Therefore it is worthwhile
to examine the higher-mass candidates with a uniform prior in $\log M$.
For the three BH candidates the likelihood with this uniform prior is
plotted as the dashed curved in the top panels of Figures
\ref{fig:likeli-mb99-22}-\ref{fig:likeli-mb98-6}.
With this uniform prior we find event MACHO-99-BLG-22 peaks at
$M\sim30\,M_\sun$, MACHO-96-BLG-5 is $M\sim10\,M_\sun$, and MACHO-98-BLG-6 is
$M\sim2\,M_\sun$.

\begin{figure}
\plotone{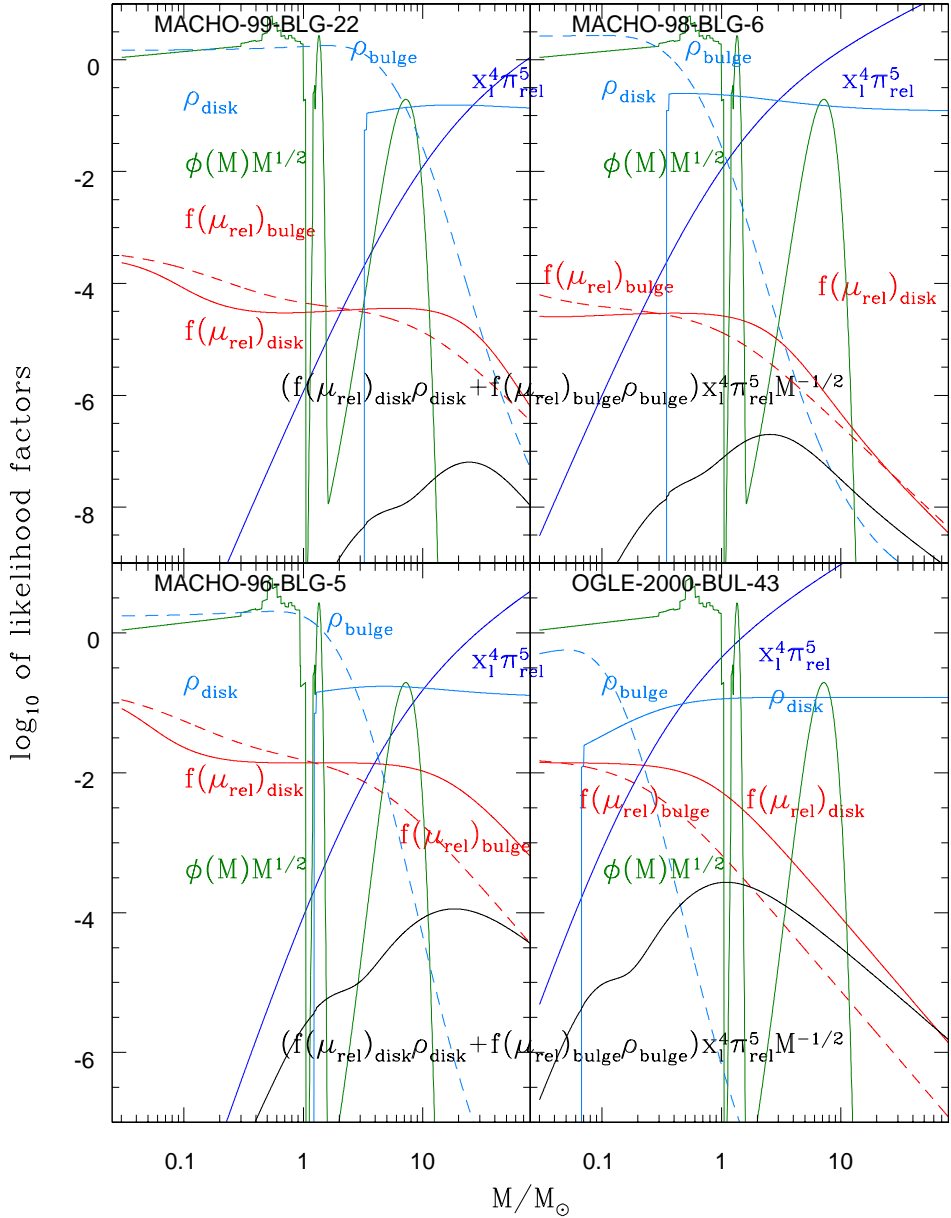}  
\caption{
How the various likelihood factors affect the
mass determination.
For simplicity, we have only plotted
these factors for a bulge source at 8 kpc while our full likelihood
calculations (\S~\ref{sec:likelihood}) are based on an integral over
source distances.
Hence, the figure only takes account of disk and
bulge lenses on our side of the Galactic center.
Shown here are the three black-hole
candidates identified in \citet{bennett02}, plus one other ``normal'' parallax
event (OGLE-2000-BUL-43). Only the lowest $\dchi$ solution is displayed for
each of these events. The black curve gives the likelihood for a uniform
log mass prior.
\label{fig:likelihoodfactors}}
\end{figure}

Figure \ref{fig:likelihoodfactors} illustrates the difficulty of interpreting 
black-hole candidates from microlens parallax information alone.
The curves labeled 
``$(f(\mu_{\rm rel})_{\rm disk}
+ f(\mu_{\rm rel})_{\rm bulge})x_l^4\pi_\rel^5\phi(M)^{-1/2}$''
represents the mass likelihood function under the assumption of a
uniform $\log M$ prior (and restricted to bulge sources at $x_s=R_0$).
Presumably a ``typical'' black hole of mass $M=7\,M_\odot$ would
have event parameters that yield a peak in this curve near $M=7\,M_\odot$.
However, MACHO-96-BLG-5 has a peak that lies above
this value and yet is ranked only as a ``marginal'' (34\%) candidate,
while MACHO-98-BLG-6 has a peak somewhat below this value and
yet is ranked as a ``weak'' (2\%) candidate.  That is, both events
have parameters that are ``typical'' of actual black holes, yet
neither is confirmed as such.  Only MACHO-99-BLG-22, whose entire
probability distribution is shifted sharply toward higher masses,
survives as a strong candidate.

The problem is that the ``typical'' event parameters generated by black
holes are also consistent with being generated by ordinary stars.
Although this occurs with somewhat lower probability for any individual
star compared to any individual black hole, there are so many more
stars than black holes (at least according to the model prior) that
the stellar explanation will usually appear ``most likely'' even
for genuine black-hole events.  It is only the events whose parameters
are indicative of such high masses that
they are virtually inconsistent with any stellar-mass object, that
will survive as ``strong candidates''.  Therefore these strong
candidates are either truly massive black-holes ($M\gg 7\,M_\odot$),
or they are ordinary, $M\sim 7\,M_\odot$, black-holes that, by chance,
happen to have extreme event parameters.

Thus, while microlens parallaxes can pick out black-hole candidates,
to really find the true black holes and to measure their mass spectrum,
it will be necessary to measure $\theta_\E$ as well as $\pi_\E$.
For black holes, which are generally expected to have large $\theta_\E$,
this may be possible using ground-based interferometers \citep{delplancke01},
but in any event can be done from space using the {\it Space Interferometry
Mission} \citep{gouldsalim99}.

Further measurements could help confirm the nature of these BH candidates.  If
we hypothesize that the lens of MACHO-96-BLG-5 is a BH of $7 M_\sun$, then we know
from the parallax fit that the lens is in the near disk.  These assumptions
yield a prediction for the proper motion of the source, which could be confirmed or
denied with {\it HST} data.  However, there are several sources of uncertainty: the
mass of the BH, the distance to the source, the velocity dispersions in the disk and
bulge, and the fact that the microlens parallax is measured to only finite precision.
This technique could be applied to other BH candidates as well.

We find this sample of parallax events are biased toward lenses that are
heavier and closer than average. This is because it is easier to detect parallax
from long time-scale events. Closer lenses give rise to 
slower $\tilde v$ which increases the time scale of the events. More massive
lenses also give rise to longer time scales.
We conclude that 5 of these events (23\%) are strongly affected by
xallarap rather than parallax.

\begin{figure}
\plotone{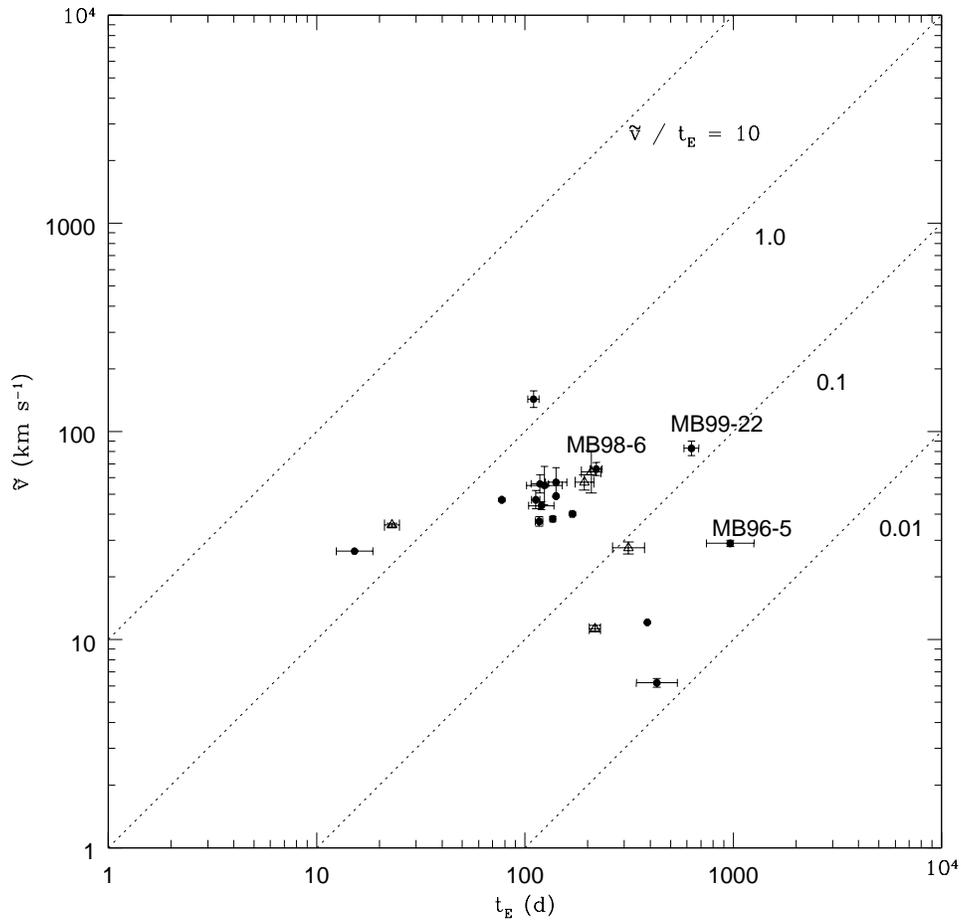}
\caption{
Projected velocity $\tilde v$ versus heliocentric timescale $t_{\rm E}$ for
the best solutions of all events in this paper.  The open triangles represent
xallarap candidates while filled circles represent parallax events.
This plot mirrors Fig. 9 of \citet{smith05} in which the four diagonal lines
indiate constant $\tilde v/t_{\rm E}=\ret/t^2_{\rm E} = 10.0,1,0.1,0.01 \kms
{\rm day}^-1$.  These represent increasingly stronger parallax deviations.
\label{fig:vtildete}}
\end{figure}

Figure \ref{fig:vtildete} shows the
projected velocity $\tilde v$ versus the heliocentric timescale $t_{\rm E}$
of the best solutions of all events from this paper. The xallarap events
(open triangles)
are not distinguished from ordinary parallax events (filled circles)
by these parameter measurements.
This figure is consistent with the predictions of Figure 9 in
\citet{smith03} which shows the distribution of simulated parallax events.

We find two additional interesting events.
Event OGLE-2003-BLG-94 is a free-floating planet candidate. However, this
is based on a $3\sigma$ finite source detection and there is no clear way
to confirm this conclusion. Event MACHO-179-A is a brown dwarf candidate
which a possibility of direct imaging in the infrared.

\acknowledgments
Work by SP and AG was supported by a NSF Grant AST-02-01266.
The paper was partly supported by the Polish MNII grant 2P03D02124.
Partial support to the OGLE project was provided with the NSF grant
AST-0204908 and NASA grant NAG5-12212. A.U. acknowledges support from
the grant ``Subsydium Profesorskie'' of the Foundation for Polish
Science.

\epsscale{.80}


\begin{deluxetable}{lllrrrrlcrrcrc}                                             
\rotate                                                                         
\tablecolumns{20}                                                               
\tablewidth{0pc}                                                                
\tablenum{1}                                                                    
\tabletypesize{\scriptsize}                                                     
\tablecaption{Microlensing Parallax Events}                                     
\tablehead{                                                                     
  \colhead{} & \multicolumn{2}{c}{Celestial} &                                  
  \multicolumn{2}{c}{Ecliptic} &                                                
  \multicolumn{2}{c}{Galactic} &                                                
  \multicolumn{2}{c}{} &                                                        
  \colhead{Standard} &                                                          
  \colhead{Best} &                                                              
  \colhead{} &                                                                  
  \colhead{Xallarap} &                                                          
  \colhead{Renomalization} \\                                                   
                                                                                
  \colhead{Event} &                                                             
  \colhead{RA} & \colhead{DEC} &                                                
  \colhead{$\lambda$} & \colhead{$\beta$} &                                     
  \colhead{$l$} & \colhead{$b$} &                                               
  \colhead{$m_{\rm base}$} &                                                    
  \colhead{points} &                                                            
  \colhead{$\Delta\chi^2$} &                                                    
  \colhead{$\chi^2$} &                                                          
  \colhead{D} &                                                                 
  \colhead{$\Delta\chi^2$} &                                                    
  \colhead{Factor}                                                              
}                                                                               
\startdata                                                                      
 MB104C\tablenotemark{1}                  & 
  18:03:34.0                              & 
 -28:00:19                                & 
 270.790 &   -4.568 &  
   2.797 &   -2.933 & 
 16.4 & 
 $ 870- 23$ & 
   1647.28 & 
    810.36 & 
 4 & 
      1.27 & 
 .8/.85 .8/.92 \\
 MB96-5\tablenotemark{1}                  & 
  18:05:02.5                              & 
 -27:42:17                                & 
 271.119 &   -4.270 &  
   3.219 &   -3.071 & 
 19.0 & 
 $ 2284- 5$ & 
   2027.74 & 
   2262.85 & 
 2 & 
     12.85 & 
 1.323  1.221  0.947
  \\
 MB96-12\tablenotemark{1,8}               & 
  18:03:53.2                              & 
 -27:57:36                                & 
 270.861 &   -4.524 &  
   2.871 &   -2.973 & 
 16.9 & 
 $ 2440- 11$ & 
  12397.19 & 
   2400.78 & 
 3 & 
      1.05 & 
 1.685  0.786  0.350  1.006  1.070
  \\
 MB98-6\tablenotemark{1}                  & 
  17:57:32.8                              & 
 -28:42:45                                & 
 268.762 &   -5.267 &  
   1.526 &   -2.132 & 
 18.5 & 
 $ 2294- 3$ & 
    557.70 & 
   2274.88 & 
 2 & 
     15.84 & 
 1.212  0.957  0.556  0.965
  \\
 MB99-1\tablenotemark{1}                  & 
  18:08:50.0                              & 
 -30:31:56                                & 
 271.917 &   -7.106 &  
   1.138 &   -5.162 & 
 18.0 & 
 $ 793- 5$ & 
   1223.97 & 
    755.99 & 
 2 & 
     15.59 & 
 1.202  0.933  0.503  0.864
  \\
 MB99-8\tablenotemark{1}                  & 
  17:56:25.2                              & 
 -29:40:31                                & 
 269.218 &   -6.237 &  
   0.569 &   -2.401 & 
 14.5 & 
 $ 1085- 4$ & 
     56.51 & 
   1063.74 & 
 2 & 
     14.72 & 
 1.596  1.299  0.934  1.591
  \\
\enddata                                                                        
\tablerefs{                                                                     
$^1$\citet{bennett02},                                                          
$^2$\citet{bond01},                                                             
$^3$\citet{smith02},                                                            
$^4$\citet{mao02,bennett02b,agol02},                                            
$^5$\citet{mao99},                                                              
$^6$\citet{soszynski01},                                                        
$^7$\citet{smithmaowoz02,smith03},                                              
$^8$\citet{afonso}}                                                             
\tablecomments{OB99-32 was also monitored by MACHO as MACHO-99-BLG-22.          
The Event names are abrivated as: MByy-xx = MACHO-yy-BLG-xx,                    
OByy-xx = OGLE-ccyy-BUL-xx, KByy-xx = MOA-ccyy-BUL-xx, OB33\_4505 = sc33\_4504, 
OC99-01 = OGLE-1999-CAR-01, EB-29 = EROS-BLG-29                                 
This table lists all 22 events examined in this paper.                          
The column $m_{\rm base}$ gives Johnson/Cousins R band for MACHO events         
and Cousins I band for all others.                                              
The column ``D'' gives the number of degenerate solutions ($\dchi2<10$).        
The degenerate solutions are in bold print in Table 2.                          
The events with ``Xallarap $\dchi$'' in bold are xallarap candidates.}          
\end{deluxetable}                                                               

\begin{deluxetable}{lrrrrrrrrrrrrrrrrrrrrr}                                     
\rotate                                                                         
\tablewidth{0pc}                                                                
\tabletypesize{\scriptsize}                                                     
\tablenum{2}                                                                    
\tablecaption{Microlensing Parallax Solutions}                                  
\tablehead{                                                                     
  \colhead{} &                                                                  
  \colhead{} &                                                                  
  \multicolumn{5}{c}{Geocentric} &                                              
  \colhead{} &                                                                  
  \colhead{} &                                                                  
  \colhead{} &                                                                  
  \colhead{} &                                                                  
  \multicolumn{4}{c}{Heliocentric} \\                                           
  \cline{3-7} \cline{10-13}                                                     
                                                                                
  \colhead{Event} &                                                             
  \colhead{$\Delta\chi^2$} &                                                    
  \colhead{$t_0$} &                                                             
  \colhead{$u_0$} &                                                             
  \colhead{$t_\E$} &                                                            
  \colhead{$\pi_{\E,N}$} &                                                      
  \colhead{$\pi_{\E,E}$} &                                                      
  \colhead{$m_{\rm source}$} &                                                  
  \colhead{$\eta_b$} &                                                          
  \colhead{$\tilde v$} &                                                        
  \colhead{$\theta$} &                                                          
  \colhead{$\tilde r_\E$} &                                                     
  \colhead{$t_\E$} \\                                                           
                                                                                
  \colhead{} &                                                                  
  \colhead{} &                                                                  
  \colhead{(JD)} &                                                              
  \colhead{} &                                                                  
  \colhead{(days)} &                                                            
  \colhead{} &                                                                  
  \colhead{} &                                                                  
  \colhead{} &                                                                  
  \colhead{} &                                                                  
  \colhead{($\kms$)} &                                                          
  \colhead{} &                                                                  
  \colhead{\au} &                                                               
  \colhead{(days)}                                                              
}                                                                               
\startdata                                                                      
  \\
 MB104C                                  
  & 
     1647.28\phantom{$^{*}$} & 
      -882.60(5)                          & 
       -0.073(1)                          & 
          179(2)                          & 
 \nodata & 
 \nodata & 
        16.76(2)                          & 
  0.27 & 
 \nodata &
 \nodata &
 \nodata &
 \nodata 
 \\*
  & 
 \bf 
        0.37\phantom{$^{*}$} & 
 \bf 
      -884.33(7)                          & 
 \bf 
       -0.083(2)                          & 
 \bf 
          169(3)                          & 
 \bf 
         0.28(4)                          & 
 \bf 
        0.163(6)                          & 
 \bf 
        16.58(2)                          & 
 \bf 
  0.15 & 
 \bf 
           47(4)                          & 
 \bf 
         52.6(9)                          & 
 \bf 
          3.1(3)                          & 
 \bf 
          112(2)                         
 \\*
  & 
 \bf 
        0.00\phantom{$^{*}$} & 
 \bf 
      -884.28(6)                          & 
 \bf 
        0.051(2)                          & 
 \bf 
          232(9)                          & 
 \bf 
         0.30(1)                          & 
 \bf 
        0.138(5)                          & 
 \bf 
        17.15(5)                          & 
 \bf 
  0.50 & 
 \bf 
           38(1)                          & 
 \bf 
         54.6(4)                          & 
 \bf 
          3.0(1)                          & 
 \bf 
          136(3)                         
 \\
\enddata                                                                        
\tablecomments{Solutions in bold are degenerate ($\dchi<10$).                   
The column $\eta_b\equiv f_b/(f_s+f_b)$                                         
gives the function of unlensed background light flux to the total               
baseline flux. The ``geocentric'' parameters are those obtained in the          
fit. The ``heliocentric'' parameters are derived from these and                 
describe the event as it would be seen from the Sun. In particular,             
$\theta$ is the angle of the lens-source relative motion, counterclockwise      
(celestial) north through east.                                                 
An asterisk after the $\dchi$ indicates that the solution has the               
background flux parameter ($f_{b,i}$ in equation [\ref{eqn:paceq}])             
fixed to zero because the unconstrained solution (listed in the                 
preceding row) has an unrealistically negative blend.}                          
\end{deluxetable}                                                               
\clearpage

\notetoeditor{The tables in table3.tex should only appear in the electronic
version. Let me know if I need to send these tables in a different format.}
  
 \begin{deluxetable}{lrrrrrrr}
 \tablecolumns{7}
 \tablewidth{0pc}
 \tablenum{ 3}
 \tabletypesize{\scriptsize}
 \tablecaption{Likelihood of MB104C                                  }
 \tablehead{
 \colhead{Lens} &
 \colhead{Source} &
 \colhead{MS} &
 \colhead{WD} &
 \colhead{NS} &
 \colhead{BH} &
 \colhead{Total}
 }
 \startdata
\sidehead{Solution 1\phantom{00000}Weight 0.25}
 \cline{1-7}
 Disk & 
 Disk & 
        1.89 &         0.64 &         0.06 &         0.00 &         2.59& \\*
 & Bulge & 
       17.43 &         9.78 &         1.12 &         0.04 &        28.37& \\*
 & Sgr & 
        0.45 &         0.19 &         0.01 &         0.00 &         0.65& \\*
 & All & 
       19.77 &        10.61 &         1.19 &         0.04 &        31.60& 
 \\* \cline{1-7} 
 Bulge & 
 Disk & 
        0.12 &         0.00 &         0.00 &         0.00 &         0.12& \\*
 & Bulge & 
        0.52 &         0.00 &         0.00 &         0.00 &         0.52& \\*
 & Sgr & 
        0.07 &         0.00 &         0.00 &         0.00 &         0.07& \\*
 & All & 
        0.70 &         0.00 &         0.00 &         0.00 &         0.70& 
 \\
\enddata
 \end{deluxetable}
 \clearpage
  
 \begin{deluxetable}{lrrrrrrr}
 \tablecolumns{7}
 \tablewidth{0pc}
 \tablenum{ 4}
 \tabletypesize{\scriptsize}
 \tablecaption{Likelihood of MB96-5                                  }
 \tablehead{
 \colhead{Lens} &
 \colhead{Source} &
 \colhead{MS} &
 \colhead{WD} &
 \colhead{NS} &
 \colhead{BH} &
 \colhead{Total}
 }
 \startdata
\sidehead{Solution 1\phantom{00000}Weight 0.36}
 \cline{1-7}
 Disk & 
 Disk & 
        0.22 &         0.35 &         3.01 &         1.81 &         5.39& \\*
 & Bulge & 
        2.41 &         1.80 &         5.56 &        23.39 &        33.17& \\*
 & Sgr & 
        0.00 &         0.00 &         0.00 &         0.32 &         0.32& \\*
 & All & 
        2.63 &         2.15 &         8.57 &        25.53 &        38.88& 
 \\* \cline{1-7} 
 Bulge & 
 Disk & 
        4.15 &         3.33 &         0.29 &         0.00 &         7.77& \\*
 & Bulge & 
       20.02 &        11.82 &         0.96 &         0.00 &        32.80& \\*
 & Sgr & 
        0.00 &         0.01 &         0.77 &         0.00 &         0.78& \\*
 & All & 
       24.16 &        15.17 &         2.02 &         0.00 &        41.35& 
 \\
\enddata
 \end{deluxetable}
 \clearpage
  
 \begin{deluxetable}{lrrrrrrr}
 \tablecolumns{7}
 \tablewidth{0pc}
 \tablenum{ 5}
 \tabletypesize{\scriptsize}
 \tablecaption{Likelihood of MB96-12                                 }
 \tablehead{
 \colhead{Lens} &
 \colhead{Source} &
 \colhead{MS} &
 \colhead{WD} &
 \colhead{NS} &
 \colhead{BH} &
 \colhead{Total}
 }
 \startdata
\sidehead{Solution 1\phantom{00000}Weight 0.38}
 \cline{1-7}
 Disk & 
 Disk & 
        2.28 &         0.99 &         0.10 &         0.00 &         3.38& \\*
 & Bulge & 
       15.36 &        10.44 &         1.65 &         0.06 &        27.52& \\*
 & Sgr & 
        2.52 &         1.20 &         0.04 &         0.00 &         3.77& \\*
 & All & 
       20.17 &        12.64 &         1.79 &         0.07 &        34.67& 
 \\* \cline{1-7} 
 Bulge & 
 Disk & 
        0.31 &         0.00 &         0.00 &         0.00 &         0.31& \\*
 & Bulge & 
        1.36 &         0.00 &         0.00 &         0.00 &         1.36& \\*
 & Sgr & 
        0.26 &         0.00 &         0.00 &         0.00 &         0.26& \\*
 & All & 
        1.92 &         0.00 &         0.00 &         0.00 &         1.92& 
 \\
\enddata
 \end{deluxetable}
 \clearpage
  
 \begin{deluxetable}{lrrrrrrr}
 \tablecolumns{7}
 \tablewidth{0pc}
 \tablenum{ 6}
 \tabletypesize{\scriptsize}
 \tablecaption{Likelihood of MB98-6                                  }
 \tablehead{
 \colhead{Lens} &
 \colhead{Source} &
 \colhead{MS} &
 \colhead{WD} &
 \colhead{NS} &
 \colhead{BH} &
 \colhead{Total}
 }
 \startdata
\sidehead{Solution 1\phantom{00000}Weight 0.97}
 \cline{1-7}
 Disk & 
 Disk & 
        4.74 &         1.48 &         1.96 &         0.13 &         8.31& \\*
 & Bulge & 
       20.90 &         9.07 &        11.08 &         1.99 &        43.03& \\*
 & Sgr & 
        0.00 &         0.00 &         0.00 &         0.00 &         0.00& \\*
 & All & 
       25.64 &        10.55 &        13.03 &         2.11 &        51.34& 
 \\* \cline{1-7} 
 Bulge & 
 Disk & 
        2.51 &         1.34 &         0.04 &         0.00 &         3.89& \\*
 & Bulge & 
       29.21 &        14.25 &         0.13 &         0.00 &        43.60& \\*
 & Sgr & 
        0.00 &         0.00 &         0.00 &         0.00 &         0.00& \\*
 & All & 
       31.72 &        15.60 &         0.17 &         0.00 &        47.49& 
 \\
\enddata
 \end{deluxetable}
 \clearpage
  
 \begin{deluxetable}{lrrrrrrr}
 \tablecolumns{7}
 \tablewidth{0pc}
 \tablenum{ 7}
 \tabletypesize{\scriptsize}
 \tablecaption{Likelihood of MB99-1                                  }
 \tablehead{
 \colhead{Lens} &
 \colhead{Source} &
 \colhead{MS} &
 \colhead{WD} &
 \colhead{NS} &
 \colhead{BH} &
 \colhead{Total}
 }
 \startdata
\sidehead{Solution 1\phantom{00000}Weight 0.45}
 \cline{1-7}
 Disk & 
 Disk & 
        3.94 &         1.62 &         0.17 &         0.01 &         5.74& \\*
 & Bulge & 
       26.66 &        14.12 &         1.36 &         0.05 &        42.19& \\*
 & Sgr & 
        0.00 &         0.00 &         0.00 &         0.00 &         0.00& \\*
 & All & 
       30.61 &        15.74 &         1.53 &         0.06 &        47.94& 
 \\* \cline{1-7} 
 Bulge & 
 Disk & 
        0.10 &         0.00 &         0.00 &         0.00 &         0.10& \\*
 & Bulge & 
        0.69 &         0.00 &         0.00 &         0.00 &         0.69& \\*
 & Sgr & 
        0.00 &         0.00 &         0.00 &         0.00 &         0.00& \\*
 & All & 
        0.79 &         0.00 &         0.00 &         0.00 &         0.79& 
 \\
\enddata
 \end{deluxetable}
 \clearpage
  
 \begin{deluxetable}{lrrrrrrr}
 \tablecolumns{7}
 \tablewidth{0pc}
 \tablenum{ 8}
 \tabletypesize{\scriptsize}
 \tablecaption{Likelihood of MB99-8                                  }
 \tablehead{
 \colhead{Lens} &
 \colhead{Source} &
 \colhead{MS} &
 \colhead{WD} &
 \colhead{NS} &
 \colhead{BH} &
 \colhead{Total}
 }
 \startdata
\sidehead{Solution 1\phantom{00000}Weight 0.57}
 \cline{1-7}
 Disk & 
 Disk & 
        0.36 &         0.11 &         0.01 &         0.00 &         0.48& \\*
 & Bulge & 
       43.48 &        20.04 &         2.89 &         0.09 &        66.50& \\*
 & Sgr & 
        0.00 &         0.00 &         0.00 &         0.00 &         0.00& \\*
 & All & 
       43.83 &        20.15 &         2.90 &         0.09 &        66.98& 
 \\* \cline{1-7} 
 Bulge & 
 Disk & 
        0.56 &         0.00 &         0.00 &         0.00 &         0.56& \\*
 & Bulge & 
        2.17 &         0.00 &         0.00 &         0.00 &         2.17& \\*
 & Sgr & 
        0.00 &         0.00 &         0.00 &         0.00 &         0.00& \\*
 & All & 
        2.73 &         0.00 &         0.00 &         0.00 &         2.73& 
 \\
\enddata
 \end{deluxetable}
 \clearpage
  
 \begin{deluxetable}{lrrrrrrr}
 \tablecolumns{7}
 \tablewidth{0pc}
 \tablenum{ 9}
 \tabletypesize{\scriptsize}
 \tablecaption{Likelihood of MB179-A                                 }
 \tablehead{
 \colhead{Lens} &
 \colhead{Source} &
 \colhead{MS} &
 \colhead{WD} &
 \colhead{NS} &
 \colhead{BH} &
 \colhead{Total}
 }
 \startdata
 Disk & 
 Disk & 
        8.37 &         0.22 &         0.01 &         0.00 &         8.60& \\*
 & Bulge & 
       88.16 &         2.13 &         0.10 &         0.00 &        90.39& \\*
 & Sgr & 
        0.99 &         0.02 &         0.00 &         0.00 &         1.01& \\*
 & All & 
       97.52 &         2.37 &         0.11 &         0.00 &       100.00& 
 \\* \cline{1-7} 
 Bulge & 
 Disk & 
        0.00 &         0.00 &         0.00 &         0.00 &         0.00& \\*
 & Bulge & 
        0.00 &         0.00 &         0.00 &         0.00 &         0.00& \\*
 & Sgr & 
        0.00 &         0.00 &         0.00 &         0.00 &         0.00& \\*
 & All & 
        0.00 &         0.00 &         0.00 &         0.00 &         0.00& 
 \\* \cline{1-7} 
 Both & 
 Disk & 
        8.37 &         0.22 &         0.01 &         0.00 &         8.60& \\*
 & Bulge & 
       88.16 &         2.13 &         0.10 &         0.00 &        90.39& \\*
 & Sgr & 
        0.99 &         0.02 &         0.00 &         0.00 &         1.01& \\*
 & All & 
       97.52 &         2.37 &         0.11 &         0.00 &       100.00& 
 \\* \cline{1-7} 
 \enddata
 \end{deluxetable}
 \clearpage
  
 \begin{deluxetable}{lrrrrrrr}
 \tablecolumns{7}
 \tablewidth{0pc}
 \tablenum{ 10}
 \tabletypesize{\scriptsize}
 \tablecaption{Likelihood of MB98-1                                  }
 \tablehead{
 \colhead{Lens} &
 \colhead{Source} &
 \colhead{MS} &
 \colhead{WD} &
 \colhead{NS} &
 \colhead{BH} &
 \colhead{Total}
 }
 \startdata
 Disk & 
 Disk & 
       10.08 &         1.24 &         0.08 &         0.00 &        11.40& \\*
 & Bulge & 
       69.82 &        15.33 &         0.91 &         0.03 &        86.09& \\*
 & Sgr & 
        1.96 &         0.18 &         0.01 &         0.00 &         2.15& \\*
 & All & 
       81.86 &        16.75 &         1.00 &         0.03 &        99.64& 
 \\* \cline{1-7} 
 Bulge & 
 Disk & 
        0.04 &         0.00 &         0.00 &         0.00 &         0.04& \\*
 & Bulge & 
        0.12 &         0.00 &         0.00 &         0.00 &         0.12& \\*
 & Sgr & 
        0.20 &         0.00 &         0.00 &         0.00 &         0.20& \\*
 & All & 
        0.36 &         0.00 &         0.00 &         0.00 &         0.36& 
 \\* \cline{1-7} 
 Both & 
 Disk & 
       10.12 &         1.24 &         0.08 &         0.00 &        11.44& \\*
 & Bulge & 
       69.94 &        15.33 &         0.91 &         0.03 &        86.21& \\*
 & Sgr & 
        2.16 &         0.18 &         0.01 &         0.00 &         2.35& \\*
 & All & 
       82.22 &        16.75 &         1.00 &         0.03 &       100.00& 
 \\* \cline{1-7} 
 \enddata
 \end{deluxetable}
 \clearpage
  
 \begin{deluxetable}{lrrrrrrr}
 \tablecolumns{7}
 \tablewidth{0pc}
 \tablenum{ 11}
 \tabletypesize{\scriptsize}
 \tablecaption{Likelihood of MB95-27                                 }
 \tablehead{
 \colhead{Lens} &
 \colhead{Source} &
 \colhead{MS} &
 \colhead{WD} &
 \colhead{NS} &
 \colhead{BH} &
 \colhead{Total}
 }
 \startdata
\sidehead{Solution 1\phantom{00000}Weight 0.32}
 \cline{1-7}
 Disk & 
 Disk & 
        2.56 &         1.66 &         0.34 &         0.03 &         4.59& \\*
 & Bulge & 
       12.66 &        10.32 &         4.05 &         0.49 &        27.53& \\*
 & Sgr & 
        1.13 &         0.60 &         0.14 &         0.01 &         1.88& \\*
 & All & 
       16.35 &        12.57 &         4.53 &         0.53 &        33.99& 
 \\* \cline{1-7} 
 Bulge & 
 Disk & 
        0.20 &         0.00 &         0.00 &         0.00 &         0.21& \\*
 & Bulge & 
        1.72 &         0.01 &         0.00 &         0.00 &         1.73& \\*
 & Sgr & 
        0.33 &         0.02 &         0.00 &         0.00 &         0.35& \\*
 & All & 
        2.26 &         0.03 &         0.00 &         0.00 &         2.29& 
 \\
\enddata
 \end{deluxetable}
 \clearpage
  
 \begin{deluxetable}{lrrrrrrr}
 \tablecolumns{7}
 \tablewidth{0pc}
 \tablenum{ 12}
 \tabletypesize{\scriptsize}
 \tablecaption{Likelihood of MB99-22                                 }
 \tablehead{
 \colhead{Lens} &
 \colhead{Source} &
 \colhead{MS} &
 \colhead{WD} &
 \colhead{NS} &
 \colhead{BH} &
 \colhead{Total}
 }
 \startdata
\sidehead{Solution 1\phantom{00000}Weight 0.68}
 \cline{1-7}
 Disk & 
 Disk & 
        0.05 &         0.03 &         0.03 &        13.87 &        13.98& \\*
 & Bulge & 
        0.00 &         0.00 &         0.02 &        31.28 &        31.30& \\*
 & Sgr & 
        0.00 &         0.00 &         0.00 &         0.00 &         0.00& \\*
 & All & 
        0.06 &         0.03 &         0.04 &        45.15 &        45.28& 
 \\* \cline{1-7} 
 Bulge & 
 Disk & 
        0.00 &         0.00 &         0.00 &         3.92 &         3.93& \\*
 & Bulge & 
        8.12 &         3.39 &         5.40 &        10.33 &        27.24& \\*
 & Sgr & 
        0.00 &         0.00 &         0.00 &         0.00 &         0.00& \\*
 & All & 
        8.12 &         3.39 &         5.40 &        14.25 &        31.16& 
 \\
\enddata
 \end{deluxetable}
 \clearpage
  
 \begin{deluxetable}{lrrrrrrr}
 \tablecolumns{7}
 \tablewidth{0pc}
 \tablenum{ 13}
 \tabletypesize{\scriptsize}
 \tablecaption{Likelihood of KB00-11                                 }
 \tablehead{
 \colhead{Lens} &
 \colhead{Source} &
 \colhead{MS} &
 \colhead{WD} &
 \colhead{NS} &
 \colhead{BH} &
 \colhead{Total}
 }
 \startdata
\sidehead{Solution 1\phantom{00000}Weight 0.44}
 \cline{1-7}
 Disk & 
 Disk & 
        3.57 &         0.39 &         0.02 &         0.00 &         3.98& \\*
 & Bulge & 
       30.41 &         5.67 &         0.28 &         0.01 &        36.37& \\*
 & Sgr & 
        2.15 &         0.11 &         0.00 &         0.00 &         2.26& \\*
 & All & 
       36.13 &         6.17 &         0.31 &         0.01 &        42.61& 
 \\* \cline{1-7} 
 Bulge & 
 Disk & 
        0.14 &         0.00 &         0.00 &         0.00 &         0.14& \\*
 & Bulge & 
        0.47 &         0.00 &         0.00 &         0.00 &         0.47& \\*
 & Sgr & 
        0.19 &         0.00 &         0.00 &         0.00 &         0.19& \\*
 & All & 
        0.81 &         0.00 &         0.00 &         0.00 &         0.81& 
 \\
\enddata
 \end{deluxetable}
 \clearpage
  
 \begin{deluxetable}{lrrrrrrr}
 \tablecolumns{7}
 \tablewidth{0pc}
 \tablenum{ 14}
 \tabletypesize{\scriptsize}
 \tablecaption{Likelihood of OB99-19                                 }
 \tablehead{
 \colhead{Lens} &
 \colhead{Source} &
 \colhead{MS} &
 \colhead{WD} &
 \colhead{NS} &
 \colhead{BH} &
 \colhead{Total}
 }
 \startdata
 Disk & 
 Disk & 
        6.26 &         2.17 &         0.21 &         0.01 &         8.66& \\*
 & Bulge & 
       54.68 &        29.41 &         3.36 &         0.15 &        87.60& \\*
 & Sgr & 
        1.43 &         0.57 &         0.04 &         0.00 &         2.04& \\*
 & All & 
       62.38 &        32.15 &         3.61 &         0.16 &        98.30& 
 \\* \cline{1-7} 
 Bulge & 
 Disk & 
        0.22 &         0.00 &         0.00 &         0.00 &         0.22& \\*
 & Bulge & 
        1.26 &         0.00 &         0.00 &         0.00 &         1.26& \\*
 & Sgr & 
        0.21 &         0.00 &         0.00 &         0.00 &         0.21& \\*
 & All & 
        1.70 &         0.00 &         0.00 &         0.00 &         1.70& 
 \\* \cline{1-7} 
 Both & 
 Disk & 
        6.48 &         2.17 &         0.21 &         0.01 &         8.88& \\*
 & Bulge & 
       55.95 &        29.41 &         3.36 &         0.15 &        88.87& \\*
 & Sgr & 
        1.64 &         0.57 &         0.04 &         0.00 &         2.25& \\*
 & All & 
       64.08 &        32.15 &         3.61 &         0.16 &       100.00& 
 \\* \cline{1-7} 
 \enddata
 \end{deluxetable}
 \clearpage
  
 \begin{deluxetable}{lrrrrrrr}
 \tablecolumns{7}
 \tablewidth{0pc}
 \tablenum{ 15}
 \tabletypesize{\scriptsize}
 \tablecaption{Likelihood of OC99-01                                 }
 \tablehead{
 \colhead{Lens} &
 \colhead{Source} &
 \colhead{MS} &
 \colhead{WD} &
 \colhead{NS} &
 \colhead{BH} &
 \colhead{Total}
 }
 \startdata
\sidehead{Solution 1\phantom{00000}Weight 0.69}
 \cline{1-7}
 Disk & 
 Disk & 
       56.49 &        25.95 &        14.61 &         0.64 &        97.69& \\*
 & Bulge & 
        0.00 &         0.00 &         0.00 &         0.00 &         0.00& \\*
 & Sgr & 
        0.00 &         0.00 &         0.00 &         0.00 &         0.00& \\*
 & All & 
       56.49 &        25.95 &        14.61 &         0.64 &        97.69& 
 \\* \cline{1-7} 
 Bulge & 
 Disk & 
        0.00 &         0.00 &         0.00 &         0.00 &         0.00& \\*
 & Bulge & 
        0.00 &         0.00 &         0.00 &         0.00 &         0.00& \\*
 & Sgr & 
        0.00 &         0.00 &         0.00 &         0.00 &         0.00& \\*
 & All & 
        0.00 &         0.00 &         0.00 &         0.00 &         0.00& 
 \\
\enddata
 \end{deluxetable}
 \clearpage
  
 \begin{deluxetable}{lrrrrrrr}
 \tablecolumns{7}
 \tablewidth{0pc}
 \tablenum{ 16}
 \tabletypesize{\scriptsize}
 \tablecaption{Likelihood of OB00-43                                 }
 \tablehead{
 \colhead{Lens} &
 \colhead{Source} &
 \colhead{MS} &
 \colhead{WD} &
 \colhead{NS} &
 \colhead{BH} &
 \colhead{Total}
 }
 \startdata
 Disk & 
 Disk & 
        4.88 &         2.24 &         0.50 &         0.05 &         7.66& \\*
 & Bulge & 
       50.54 &        24.12 &         6.79 &         0.50 &        81.95& \\*
 & Sgr & 
        4.86 &         2.74 &         0.26 &         0.01 &         7.86& \\*
 & All & 
       60.28 &        29.09 &         7.55 &         0.56 &        97.48& 
 \\* \cline{1-7} 
 Bulge & 
 Disk & 
        0.09 &         0.00 &         0.00 &         0.00 &         0.09& \\*
 & Bulge & 
        1.91 &         0.00 &         0.00 &         0.00 &         1.91& \\*
 & Sgr & 
        0.51 &         0.01 &         0.00 &         0.00 &         0.52& \\*
 & All & 
        2.51 &         0.01 &         0.00 &         0.00 &         2.52& 
 \\* \cline{1-7} 
 Both & 
 Disk & 
        4.97 &         2.24 &         0.50 &         0.05 &         7.76& \\*
 & Bulge & 
       52.45 &        24.12 &         6.79 &         0.50 &        83.86& \\*
 & Sgr & 
        5.37 &         2.74 &         0.26 &         0.01 &         8.38& \\*
 & All & 
       62.79 &        29.10 &         7.55 &         0.56 &       100.00& 
 \\* \cline{1-7} 
 \enddata
 \end{deluxetable}
 \clearpage
  
 \begin{deluxetable}{lrrrrrrr}
 \tablecolumns{7}
 \tablewidth{0pc}
 \tablenum{ 17}
 \tabletypesize{\scriptsize}
 \tablecaption{Likelihood of OB33\_4505                              }
 \tablehead{
 \colhead{Lens} &
 \colhead{Source} &
 \colhead{MS} &
 \colhead{WD} &
 \colhead{NS} &
 \colhead{BH} &
 \colhead{Total}
 }
 \startdata
\sidehead{Solution 1\phantom{00000}Weight 0.02}
 \cline{1-7}
 Disk & 
 Disk & 
        0.35 &         0.22 &         0.03 &         0.00 &         0.60& \\*
 & Bulge & 
        3.44 &         3.47 &         0.70 &         0.03 &         7.63& \\*
 & Sgr & 
        0.01 &         0.00 &         0.00 &         0.00 &         0.02& \\*
 & All & 
        3.79 &         3.70 &         0.73 &         0.03 &         8.25& 
 \\* \cline{1-7} 
 Bulge & 
 Disk & 
        0.04 &         0.00 &         0.00 &         0.00 &         0.04& \\*
 & Bulge & 
        0.31 &         0.00 &         0.00 &         0.00 &         0.31& \\*
 & Sgr & 
        0.03 &         0.00 &         0.00 &         0.00 &         0.03& \\*
 & All & 
        0.38 &         0.00 &         0.00 &         0.00 &         0.38& 
 \\
\enddata
 \end{deluxetable}
 \clearpage
  
 \begin{deluxetable}{lrrrrrrr}
 \tablecolumns{7}
 \tablewidth{0pc}
 \tablenum{ 18}
 \tabletypesize{\scriptsize}
 \tablecaption{Likelihood of OB02-100                                }
 \tablehead{
 \colhead{Lens} &
 \colhead{Source} &
 \colhead{MS} &
 \colhead{WD} &
 \colhead{NS} &
 \colhead{BH} &
 \colhead{Total}
 }
 \startdata
\sidehead{Solution 1\phantom{00000}Weight 0.42}
 \cline{1-7}
 Disk & 
 Disk & 
        3.56 &         0.15 &         0.01 &         0.00 &         3.71& \\*
 & Bulge & 
       34.28 &         0.91 &         0.04 &         0.00 &        35.23& \\*
 & Sgr & 
        0.43 &         0.01 &         0.00 &         0.00 &         0.44& \\*
 & All & 
       38.27 &         1.06 &         0.05 &         0.00 &        39.38& 
 \\* \cline{1-7} 
 Bulge & 
 Disk & 
        0.00 &         0.00 &         0.00 &         0.00 &         0.00& \\*
 & Bulge & 
        0.00 &         0.00 &         0.00 &         0.00 &         0.00& \\*
 & Sgr & 
        0.00 &         0.00 &         0.00 &         0.00 &         0.00& \\*
 & All & 
        0.00 &         0.00 &         0.00 &         0.00 &         0.00& 
 \\
\enddata
 \end{deluxetable}
 \clearpage
  
 \begin{deluxetable}{lrrrrrrr}
 \tablecolumns{7}
 \tablewidth{0pc}
 \tablenum{ 19}
 \tabletypesize{\scriptsize}
 \tablecaption{Likelihood of OB02-334                                }
 \tablehead{
 \colhead{Lens} &
 \colhead{Source} &
 \colhead{MS} &
 \colhead{WD} &
 \colhead{NS} &
 \colhead{BH} &
 \colhead{Total}
 }
 \startdata
 Disk & 
 Disk & 
        2.71 &         1.40 &         0.22 &         0.02 &         4.35& \\*
 & Bulge & 
       52.99 &        32.78 &         6.90 &         0.30 &        92.97& \\*
 & Sgr & 
        0.32 &         0.17 &         0.04 &         0.00 &         0.52& \\*
 & All & 
       56.02 &        34.34 &         7.17 &         0.31 &        97.84& 
 \\* \cline{1-7} 
 Bulge & 
 Disk & 
        0.15 &         0.00 &         0.00 &         0.00 &         0.15& \\*
 & Bulge & 
        1.81 &         0.00 &         0.00 &         0.00 &         1.81& \\*
 & Sgr & 
        0.19 &         0.00 &         0.00 &         0.00 &         0.20& \\*
 & All & 
        2.16 &         0.00 &         0.00 &         0.00 &         2.16& 
 \\* \cline{1-7} 
 Both & 
 Disk & 
        2.86 &         1.40 &         0.22 &         0.02 &         4.50& \\*
 & Bulge & 
       54.80 &        32.78 &         6.90 &         0.30 &        94.78& \\*
 & Sgr & 
        0.51 &         0.17 &         0.04 &         0.00 &         0.72& \\*
 & All & 
       58.17 &        34.35 &         7.17 &         0.31 &       100.00& 
 \\* \cline{1-7} 
 \enddata
 \end{deluxetable}
 \clearpage
  
 \begin{deluxetable}{lrrrrrrr}
 \tablecolumns{7}
 \tablewidth{0pc}
 \tablenum{ 20}
 \tabletypesize{\scriptsize}
 \tablecaption{Likelihood of OB02-61                                 }
 \tablehead{
 \colhead{Lens} &
 \colhead{Source} &
 \colhead{MS} &
 \colhead{WD} &
 \colhead{NS} &
 \colhead{BH} &
 \colhead{Total}
 }
 \startdata
\sidehead{Solution 1\phantom{00000}Weight 0.24}
 \cline{1-7}
 Disk & 
 Disk & 
        3.38 &         1.15 &         0.91 &         0.08 &         5.52& \\*
 & Bulge & 
        7.31 &         3.25 &         3.81 &         1.29 &        15.66& \\*
 & Sgr & 
        0.00 &         0.00 &         0.02 &         0.01 &         0.03& \\*
 & All & 
       10.69 &         4.40 &         4.75 &         1.37 &        21.21& 
 \\* \cline{1-7} 
 Bulge & 
 Disk & 
        2.16 &         0.70 &         0.00 &         0.00 &         2.86& \\*
 & Bulge & 
        2.92 &         1.18 &         0.00 &         0.00 &         4.10& \\*
 & Sgr & 
        0.25 &         0.06 &         0.00 &         0.00 &         0.31& \\*
 & All & 
        5.33 &         1.94 &         0.01 &         0.00 &         7.28& 
 \\
\enddata
 \end{deluxetable}
 \clearpage
  
 \begin{deluxetable}{lrrrrrrr}
 \tablecolumns{7}
 \tablewidth{0pc}
 \tablenum{ 21}
 \tabletypesize{\scriptsize}
 \tablecaption{Likelihood of OB03-188                                }
 \tablehead{
 \colhead{Lens} &
 \colhead{Source} &
 \colhead{MS} &
 \colhead{WD} &
 \colhead{NS} &
 \colhead{BH} &
 \colhead{Total}
 }
 \startdata
\sidehead{Solution 1\phantom{00000}Weight 0.02}
 \cline{1-7}
 Disk & 
 Disk & 
        0.21 &         0.09 &         0.01 &         0.00 &         0.31& \\*
 & Bulge & 
        2.41 &         1.26 &         0.19 &         0.01 &         3.87& \\*
 & Sgr & 
        0.26 &         0.12 &         0.01 &         0.00 &         0.39& \\*
 & All & 
        2.88 &         1.47 &         0.20 &         0.01 &         4.57& 
 \\* \cline{1-7} 
 Bulge & 
 Disk & 
        0.01 &         0.00 &         0.00 &         0.00 &         0.01& \\*
 & Bulge & 
        0.13 &         0.00 &         0.00 &         0.00 &         0.13& \\*
 & Sgr & 
        0.03 &         0.00 &         0.00 &         0.00 &         0.03& \\*
 & All & 
        0.17 &         0.00 &         0.00 &         0.00 &         0.17& 
 \\
\enddata
 \end{deluxetable}
 \clearpage
  
 \begin{deluxetable}{lrrrrrrr}
 \tablecolumns{7}
 \tablewidth{0pc}
 \tablenum{ 22}
 \tabletypesize{\scriptsize}
 \tablecaption{Likelihood of OB03-32                                 }
 \tablehead{
 \colhead{Lens} &
 \colhead{Source} &
 \colhead{MS} &
 \colhead{WD} &
 \colhead{NS} &
 \colhead{BH} &
 \colhead{Total}
 }
 \startdata
\sidehead{Solution 1\phantom{00000}Weight 0.99}
 \cline{1-7}
 Disk & 
 Disk & 
        5.18 &         1.34 &         0.11 &         0.00 &         6.63& \\*
 & Bulge & 
       62.27 &        26.01 &         2.07 &         0.07 &        90.43& \\*
 & Sgr & 
        0.88 &         0.29 &         0.02 &         0.00 &         1.19& \\*
 & All & 
       68.34 &        27.64 &         2.20 &         0.07 &        98.25& 
 \\* \cline{1-7} 
 Bulge & 
 Disk & 
        0.24 &         0.00 &         0.00 &         0.00 &         0.24& \\*
 & Bulge & 
        1.10 &         0.00 &         0.00 &         0.00 &         1.10& \\*
 & Sgr & 
        0.17 &         0.00 &         0.00 &         0.00 &         0.17& \\*
 & All & 
        1.51 &         0.00 &         0.00 &         0.00 &         1.51& 
 \\
\enddata
 \end{deluxetable}
 \clearpage
  
 \begin{deluxetable}{lrrrrrrr}
 \tablecolumns{7}
 \tablewidth{0pc}
 \tablenum{ 23}
 \tabletypesize{\scriptsize}
 \tablecaption{Likelihood of OB03-84                                 }
 \tablehead{
 \colhead{Lens} &
 \colhead{Source} &
 \colhead{MS} &
 \colhead{WD} &
 \colhead{NS} &
 \colhead{BH} &
 \colhead{Total}
 }
 \startdata
\sidehead{Solution 1\phantom{00000}Weight 0.02}
 \cline{1-7}
 Disk & 
 Disk & 
        0.09 &         0.03 &         0.00 &         0.00 &         0.13& \\*
 & Bulge & 
        1.01 &         0.56 &         0.04 &         0.00 &         1.61& \\*
 & Sgr & 
        0.04 &         0.01 &         0.00 &         0.00 &         0.04& \\*
 & All & 
        1.14 &         0.60 &         0.04 &         0.00 &         1.78& 
 \\* \cline{1-7} 
 Bulge & 
 Disk & 
        0.00 &         0.00 &         0.00 &         0.00 &         0.00& \\*
 & Bulge & 
        0.01 &         0.00 &         0.00 &         0.00 &         0.01& \\*
 & Sgr & 
        0.00 &         0.00 &         0.00 &         0.00 &         0.00& \\*
 & All & 
        0.01 &         0.00 &         0.00 &         0.00 &         0.01& 
 \\
\enddata
 \end{deluxetable}
 \clearpage
  
 \begin{deluxetable}{lrrrrrrr}
 \tablecolumns{7}
 \tablewidth{0pc}
 \tablenum{ 24}
 \tabletypesize{\scriptsize}
 \tablecaption{Likelihood of EB-29                                   }
 \tablehead{
 \colhead{Lens} &
 \colhead{Source} &
 \colhead{MS} &
 \colhead{WD} &
 \colhead{NS} &
 \colhead{BH} &
 \colhead{Total}
 }
 \startdata
\sidehead{Solution 1\phantom{00000}Weight 0.19}
 \cline{1-7}
 Disk & 
 Disk & 
        1.03 &         0.45 &         0.08 &         0.01 &         1.56& \\*
 & Bulge & 
        7.83 &         3.64 &         0.98 &         0.09 &        12.55& \\*
 & Sgr & 
        0.57 &         0.31 &         0.03 &         0.00 &         0.92& \\*
 & All & 
        9.43 &         4.40 &         1.10 &         0.10 &        15.03& 
 \\* \cline{1-7} 
 Bulge & 
 Disk & 
        0.04 &         0.00 &         0.00 &         0.00 &         0.04& \\*
 & Bulge & 
        0.48 &         0.00 &         0.00 &         0.00 &         0.48& \\*
 & Sgr & 
        0.12 &         0.00 &         0.00 &         0.00 &         0.12& \\*
 & All & 
        0.63 &         0.00 &         0.00 &         0.00 &         0.63& 
 \\
\enddata
 \end{deluxetable}
 \clearpage

\end{document}